\documentclass[aps,prd,twocolumn,preprintnumbers,nofootinbib,superscriptaddress,amsmath]{revtex4-2}
\usepackage{graphicx}
\usepackage{amsmath}
\usepackage{amssymb}
\usepackage{amsfonts}
\usepackage{hyperref}
\usepackage{url}
\usepackage{color}
\usepackage{bm}
\usepackage{array}
\usepackage{comment}

\newcommand{\be}{\begin{equation}}
\newcommand{\ee}{\end{equation}}
\newcommand{\bea}{\begin{eqnarray}}
\newcommand{\eea}{\end{eqnarray}}
\newcommand{\nn}{\nonumber}
\newcommand{\dd}{\mathrm{d}}
\newcommand{\Mp}{M_{\mathrm{Pl}}}

\renewcommand{\[}{\begin{equation}}
\renewcommand{\]}{\end{equation}}

\def\lcdm{$\Lambda$CDM }

\usepackage{array}
\makeatother

\begin{document}

\preprint{IFT-UAM/CSIC-22-141}

\title{Machine learning cosmic inflation}

\author{Ahana Kamerkar}
\email{ahana.kamerkar19@imperial.ac.uk}
\affiliation{Department of Theoretical Physics, Imperial College, London, SW7 2AZ, UK}
\affiliation{Instituto de F\'isica Te\'orica UAM-CSIC, Universidad Auton\'oma de Madrid, Cantoblanco, 28049 Madrid, Spain}

\author{Savvas Nesseris}
\email{savvas.nesseris@csic.es}
\affiliation{Instituto de F\'isica Te\'orica UAM-CSIC, Universidad Auton\'oma de Madrid, Cantoblanco, 28049 Madrid, Spain}

\author{Lucas Pinol}
\email{lucas.pinol@ift.csic.es}
\affiliation{Instituto de F\'isica Te\'orica UAM-CSIC, Universidad Auton\'oma de Madrid, Cantoblanco, 28049 Madrid, Spain}

\date{\today}

\begin{abstract}
We present a machine-learning approach, based on the genetic algorithms (GA), that can be used to reconstruct the inflationary potential directly from cosmological data.
We create a pipeline consisting of the GA, a primordial code and a Boltzmann code used to calculate the theoretical predictions, and Cosmic Microwave Background (CMB) data.
As a proof of concept, we apply our methodology to the Planck CMB data and explore the functional space of single-field inflationary potentials in a non-parametric, yet analytical way.
We show that the algorithm easily improves upon the vanilla model of quadratic inflation and proposes slow-roll potentials better suited to the data, while we confirm the robustness of the Starobinsky inflation model (and other small-field models).
Moreover, using unbinned CMB data, we perform a first concrete application of the GA by searching for oscillatory features in the potential in an agnostic way, and find very significant improvements upon the best featureless potentials, $\Delta \chi^2 < -20$.
These encouraging preliminary results motivate the search for resonant features in the primordial power spectrum with a multimodal distribution of frequencies.
We stress that our pipeline is modular and can easily be extended to other CMB data sets and inflationary scenarios, like multifield inflation or theories with higher-order derivatives.
\end{abstract}

\maketitle

\section{Introduction \label{sec:intro}}
Inflation is arguably the most successful scenario of the very early Universe.
This cosmological epoch of accelerated expansion enables not only to explain the seemingly acausal large-scale correlations in the Cosmic Microwave Background (CMB), but also to predict with a high degree of precision the statistics of anisotropies at all observable scales in this fossil light.
In this work, we focus on the possibility that inflation is due to the presence of a scalar field with a potential energy much larger than its kinetic energy, and with canonical kinetic terms.
These so-called single-field slow-roll models of chaotic inflation proved to be tremendously successful at explaining the statistics of inhomogeneities in the CMB, and at providing a natural end to inflation. 

However, with data of increasing precision being gathered by CMB experiments, the region of observationally allowed single-field models has dramatically narrowed down in the past years.
In particular, the Planck data (see Ref.~\cite{Planck:2018nkj} about the Planck legacy) now severely constrains single-field scalar potentials, and has already excluded several interesting possibilities proposed in the past.
It is therefore a natural question to ask: which single-field scenario results in CMB anisotropies that best match the current data?

\vskip 4pt
Of course, this question has already been addressed in the past, both by the Planck community and by independent teams.
The traditional approach consists in performing a model-by-model comparison by computing observables for several single-field slow-roll potentials, a most famous example of such study being the \textit{Encyclopaedia inflationaris}~\cite{Martin:2013tda}.
Using recent CMB data, the Planck team proposed a less exhaustive yet popular study of which of those potentials were still compatible with more precise observational constraints~\cite{Planck:2018jri}.
In these works, by ``observational constraints" is actually meant the compressed information contained in the values of the parameters $(n_s,r)$ describing the primordial power spectra, $\mathcal{P}_\zeta(k) = A_s\,(k/k_\star)^{n_s-1}$ in the scalar sector and $\mathcal{P}_\gamma(k_\star)=r \, A_s$ in the tensor sector, once $A_s$ has been fixed to the observed value at the pivot scale $k_\star$ (either $ 0.002 \,\mathrm{Mpc}^{-1}$ or $ 0.05 \,\mathrm{Mpc}^{-1}$).
Features beyond scale-invariance in the primordial scalar power spectrum have also been investigated in light of concrete models producing either resonant or sharp features (see, e.g., the review~\cite{Slosar:2019gvt}), and performing a Monte Carlo exploration of parameter space to find the best-fits to current data beyond the $(n_s,r)$ plane.

But other studies bypass the model approach by reconstructing the inflationary potential in the observable window in an agnostic way.
This is the case, e.g., of a dedicated section in Ref.~\cite{Planck:2018jri} where parameters of the Taylor expansion of the inflationary potential are inferred from data, and where a free-form reconstruction using cubic splines is also done (see Ref.~\cite{Handley:2019fll} for more details about this non-parametric but non-analytic method, as well as Ref.~\cite{Hazra:2014jwa} for an earlier work).

The novelty of our approach resides in the methodology for proposing new scalar potentials.
Using a particular machine learning (ML) technique, the so-called genetic algorithms (GAs), we allow a population of potential seeds to mutate generation after generation into random directions in an abstract functional space.
At each generation, only those members of the population that lead to CMB statistics that best match the data survive and are allowed to evolve into the next generation.
In this way, the GA explores large regions of possible potentials, while gradually improving the fit to the data until reaching a minimum.
This final potential learnt from the algorithm is then the ``best" candidate potential for single-field slow-roll inflation with canonical kinetic terms.
The main advantage of ML methods, such as the one employed in this work, is that they allow for a bottom-up reconstruction of the inflationary potential $V(\phi)$ solely from the data. Thus, they bypass possible theoretical biases and make it possible to look for features in the CMB data that may not otherwise be easily detected via more traditional approaches (see, e.g. Ref.\cite{Arjona:2020kco} in the context of dark energy).

Let us comment that a few works have already tried to embrace a machine-learning approach to infer the inflationary potential preferred by data.
This is the case of  Ref.~\cite{Rudelius:2018yqi} where the parameters of the Taylor expansion of the potential are found as weights in a single-layer neural network.
Also, Ref.~\cite{Abel:2022nje} performs sextic polynomial reconstructions where coefficients are found using a GA approach.
Both approaches are therefore parametric.
Moreover, the ``data" used in both studies is simply the values of the two parameters $(n_s,r)$, thus restricting \textit{de facto} the analysis to slow-roll inflation leading to power-law power spectra.

In this work, we take full advantage of the GA approach to look in a model-independent and analytical way for inflationary potentials preferred by data.
Importantly, we do not restrict to the two parameters $(n_s,r)$ and consider the full Planck 2018 data made of $TT$, $TE$ and $EE$ (as well as lensing) power spectra. 
This allows us to perform the first proof of concept of machine-learning inference of the inflationary potential preferred by CMB data, in a non-parametric yet analytical way.
In particular, our approach enables for a model-independent search for primordial features directly at the level of the inflationary potential.

\vskip 4pt
The structure of our paper is as follows: in Sec.~\ref{sec:theory} we quickly review the theory of inflation assuming a single scalar field $\phi$, in Sec.~\ref{sec:ga} we give a brief summary of the GA approach and how it works, while in Sec.~\ref{sec:methods} we present the data used in our analysis and our methodology. Finally, in Sec.~\ref{sec:results} we present the results of our analysis including slow-roll potentials and oscillatory features.
In Sec.~\ref{sec:conclusions} we summarize our conclusions.

\section{Theory \label{sec:theory}}

\paragraph*{Model and background dynamics.}
We assume that the matter content in the early Universe is composed of a single scalar field $\phi$, the inflaton, with canonical kinetic terms and a scalar potential $V(\phi)$, and with a minimal coupling to gravity.
It is then straightforward to derive the two Friedmann equations relating the homogeneous part of the spacetime geometry to the matter field, and the Klein-Gordon equation ruling the evolution of the scalar field itself:
\begin{align}
    H^2&=\frac{1}{3 \Mp^2}\left[\frac12\dot{\phi}^2+V(\phi)\right],\\
\dot{H}&=-\frac{1}{2\Mp^2}\dot{\phi}^2,\\
\ddot{\phi}&+3H\dot{\phi}+\frac{\dd V(\phi)}{\dd\phi}=0,
\end{align}
where $\Mp\equiv \left(8\pi\,G_N\right)^{-1/2}$ is the reduced Planck mass, $H=\dot{a}/a$ is the rate of expansion of the Universe, $a(t)$ being the scale factor and a dot meaning a derivative with respect to cosmic time $t$.
This system of differential equations for spatially homogeneous functions is closed and can be solved, either numerically or analytically in some approximation scheme.

Often, the slow-roll hypothesis of a very flat potential leading to an overdamped regime with negligible acceleration of the scalar field is assumed, which enables for an analytical resolution of the background equations in terms of the so-called slow-roll parameters, $\epsilon=-\dot{H}/H^2$ and $\eta=\dot{\epsilon}/(H \epsilon)$, which are both small in this regime.
Although the slow-roll regime is not strictly needed to apply (we shall not use such approximation in the following), it ought to be said that a successful inflationary scenario needs to provide a sufficiently long period of accelerated expansion to explain the correlations of very large scales in the CMB, which requires any way to have a relatively flat scalar potential for a sufficiently large range of $\phi$.

\paragraph*{Dynamics of linear fluctuations.}
The success of inflation lies in its predictability power for the statistics of inhomogeneities.
In order to describe them, we need to appeal to cosmological perturbation theory, whence both the spacetime geometry and the scalar field are expanded into linear fluctuations around the homogeneous, time-dependent, background presented in the previous paragraph.
The propagating degrees of freedom during single-field inflation consist in the two polarisation modes $\gamma^\lambda\,, \,\, \lambda\in\{+,\times\}\,,$ of gravitational waves and the primordial curvature perturbation $\zeta$ which is the gauge-invariant combination of the scalar field $\phi$ and spacetime metric perturbations.
They can be related to canonically normalised variables, called the Sasaki-Mukhanov variables, as $v^\lambda= (a  \Mp /2) \gamma^\lambda$ and $v=z\,\zeta$ with $z=a\sqrt{2\epsilon}\Mp$.
In Fourier space, they verify the following equations of motion, where a prime ${}\prime$ represents a derivative with respect to conformal time $\tau$ (related to cosmic time by $\dd t = a \dd \tau$):
\begin{align}
    v_k^{\lambda\prime\prime}+\left(k^2-\frac{a^{\prime\prime}}{a}\right) v_k^\lambda &= 0 \,, \\
    v_k^{\prime\prime}+\left(k^2-\frac{z^{\prime\prime}}{z}\right) v_k &= 0  \,.
\end{align}
Initial conditions for the Sasaki-Mukhanov variables are found upon canonical quantisation of the system, and identification of the so-called Bunch-Davies vacuum for every $k$-mode when they are still deep inside the comoving Hubble radius, $k \gg aH$.
The differential equations can then be solved, and one may define the dimensionless primordial tensor and scalar spectra at the end of inflation as:
\begin{align}
    \mathcal{P}_\gamma(k) &= \frac{k^3}{2\pi^2} \underset{- k\tau \rightarrow 0}{\mathrm{lim}} \sum_{\lambda=+,\times}\left(\frac{2}{a\Mp}\right)^2\left|v_k^\lambda\right|^2 \,, \\
    \mathcal{P}_\zeta(k) &= \frac{k^3}{2\pi^2} \underset{- k\tau \rightarrow 0}{\mathrm{lim}} \frac{\left|v_k\right|^2}{z^2} \,.   
\end{align}
It is customary to define the amplitudes $A_{t,s}$ and tilts $n_{t,s}$ of tensors and scalars, as $\mathcal{P}_\gamma(k)=A_t\,\left(k/k_\star\right)^{n_t}\,,\,\, \mathcal{P}_\zeta(k)=A_s\,\left(k/k_\star\right)^{n_s-1}\,,$ with $k_\star$ a reference scale called the pivot scale, which can be taken to be $k_\star=0.05\,\mathrm{Mpc}^{-1}$ in the Planck data for example.
Using this parameterisation, the latest Planck data analysis measured accurately the scalar sector, $\ln(10^{10}\,A_s) = 3.044 \pm 0.014$, $n_s=0.9649 \pm 0.0042$, and put constraints (at $95\%$ confidence level upper limit) on the maximum amount of tensor modes, $r_{0.002}=A_t/A_s < 0.10 $, therefore constraining the inflationary models to the ones with a potential correctly predicting those effective parameters~\cite{Planck:2018jri}.
However, in order to allow deviations from the simple power-law scale dependence,\textit{ we will instead not use such a parameterisation}\footnote{Some data analyses do allow for a running of the scalar spectral index $n_s(k)$, in the form of $\alpha_s =\left. \dd \, n_s(k) / (\dd \, \mathrm{ln} k) \right|_{k_\star}$, current constraints being $\alpha_s = 0.0045\pm 0.0067$.
Note however that this running cannot encompass complicated features in the primordial scalar power spectrum.
We will not use the running in our analysis neither.}, and simply use the full information contained in $\mathcal{P}_{\gamma,\zeta}(k)$ to compute the statistics of CMB anisotropies, and constrain the inflationary models.
\paragraph*{CMB anisotropies.} The angular power spectra of temperature and polarisation anisotropies, the $C_\ell$, are computed from the primordial fluctuations by means of a convolution over so-called transfer functions.
For example, scalar fluctuations source temperature anisotropies as: 
\begin{equation}
    C_\ell^{TT} \supset 4\pi \int \frac{\mathrm{d} k}{k} \mathcal{P}_\zeta(k) \left[\Delta_\ell^T(k)\right]^2\,,
\end{equation}
where $\Delta_\ell^T(k)$ is the corresponding transfer function, and similar relations hold for other kinds of anisotropies, including cross-correlations.
Computing rigorously the transfer functions is not an easy task -- it is precisely the aim of the so-called Einstein-Boltzmann codes to solve for them in order to calculate the angular power spectra.
For this, the hierarchy of Boltzmann equations for the matter content in the early Universe (photons, baryons, dark matter, neutrinos, etc.) is truncated at a given order.
They are solved in the expanding background and taking into account metric perturbations, which are consistently evolved via the Friedmann and constraint equations. Once the dynamics of the matter content has been solved, the transfer functions can be computed with the line-of-sight formalism, where they are found upon integration along this line of sight. See the classic reference~\cite{Ma:1995ey} for more details.

\begin{figure*}[!t]
    \centering
    \includegraphics[width=0.85\textwidth]{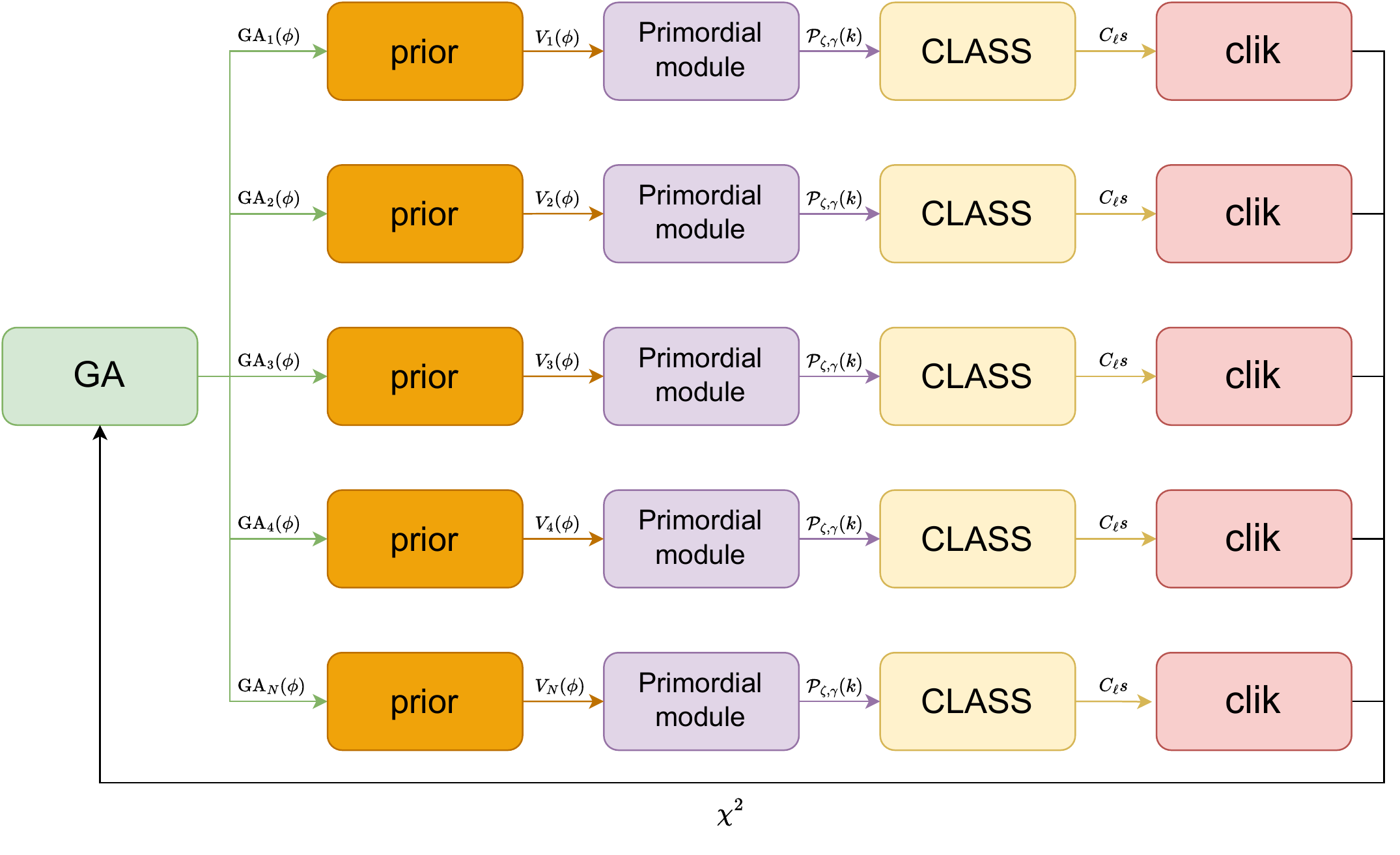}
    \caption{A flowchart briefly summarizing our methodology: First the GA creates a set of random functions $\mathrm{GA}_i(\phi)$ of the scalar field $\phi$, where $i\in\{1,\ldots,N\}$ and $N$ is the population size. Each of these functions is then combined with the prior in order to create a potential $V_i(\phi)$, that is subsequently fed to the ``Primordial module" that calculates the inflationary predictions. The latter are passed on to a Boltzmann code, in our case  \texttt{CLASS}, in order to calculate the theoretical CMB predictions of each potential, which are in turn passed to the Planck likelihood \texttt{clik} (including the temperature, polarization and lensing spectra) that returns the log-likelihood, i.e. the $\chi^2$, for each potential. In our analysis the primordial package used is the one included in \texttt{CLASS}, but any other one could also be used. Finally, this process is repeated hundreds of times to ensure convergence.}
    \label{fig:flow}
\end{figure*}

In this work, we consider the effect of both primordial scalar and tensor modes on CMB statistics.
However, we will use public Planck likelihoods which only contain information about the $TT$, $TE$ and $EE$ anisotropies of the CMB (as well as lensing).
In principle, constraints on primordial $B$ modes from other experiments such as BICEP-Keck also contain a handful quantity of information on the inflationary epoch, in particular via the stringent constrains that they provide on the maximum amount of primordial tensor modes~\cite{BICEP:2021xfz}.
Although Planck 18 data does not include primordial $B$ modes, it already provides a weak constraint on $r_{0.002}<0.10$.
As this work is a proof of concept, we leave for future work the inclusion of other likelihoods from other experiments, a task that should be straightforward to implement given the modularity or our approach explained in Sec.~\ref{sec:methods} and represented in Fig.~\ref{fig:flow}.

\section{The genetic algorithms \label{sec:ga}}
\paragraph*{Overview.}
In this section we present in detail our implementation of the GA.
GAs have been used extensively in the past in various applications in cosmology, e.g. see  Refs.~\cite{Bogdanos:2009ib,Nesseris:2010ep, Nesseris:2012tt,Nesseris:2013bia,Sapone:2014nna,Arjona:2020doi,Arjona:2020kco,Arjona:2019fwb}, also in forecasts for forthcoming large scale structure surveys \cite{EUCLID:2020syl, Euclid:2021cfn, Euclid:2021frk}, but also in other areas such as particle physics \cite{Abel:2018ekz,Allanach:2004my,Akrami:2009hp} and astronomy and astrophysics \cite{wahde2001determination,Rajpaul:2012wu,Ho:2019zap}. Applications in fields outside physics include computational science, economics, medicine and engineering \cite{affenzeller2009genetic,sivanandam2008genetic}. Finally, there have also been several different symbolic regression approaches such as the ones described in Refs.~\cite{Udrescu:2019mnk,Setyawati:2019xzw,vaddireddy2019feature,Liao:2019qoc,Belgacem:2019zzu,Li:2019kdj,Bernardini:2019bmd,Gomez-Valent:2019lny,Bartlett:2022kyi}.

\begin{figure}[!t]
\centering
\includegraphics[width = 0.5\textwidth]{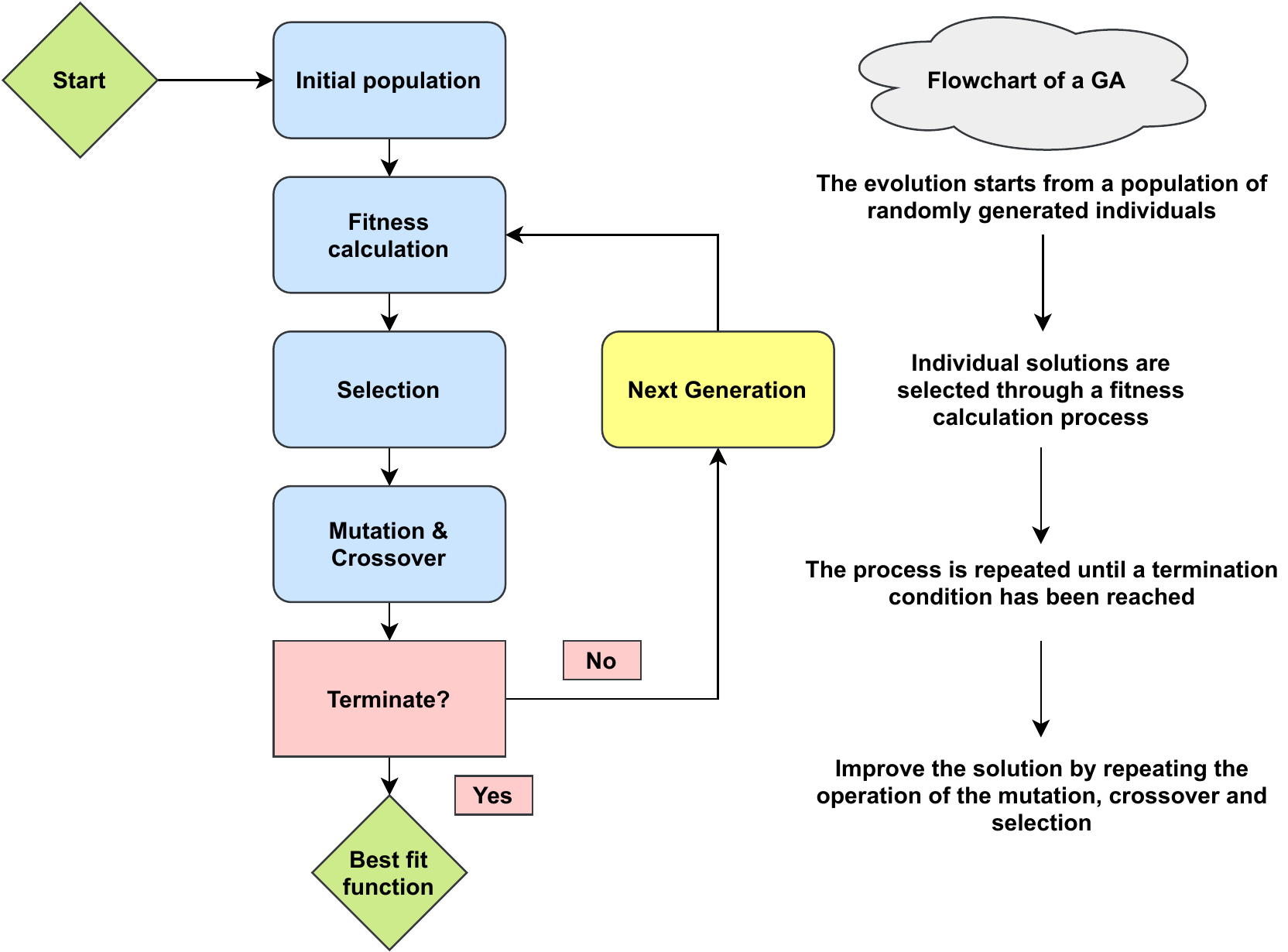}
\caption{Flowchart of a typical run of a genetic algorithm (from Ref.~\cite{Arjona:2020axn}). \label{fig:GA_flowchart}}
\end{figure}

The GA is a particular ML approach that performs an unsupervised symbolic regression of data, thus it is ideal for non-parametric analytic reconstructions that describe the data via one or more variables, such as the scalar field $\phi$. 
The fact that the GA performs an unsupervised regression implies that the algorithm extracts features and patterns from the data without external help or labelling. On the contrary, supervised machine-learning algorithms train on labeled datasets in order to provide good predictions. In Fig.~\ref{fig:GA_flowchart} we also show a flowchart that summarizes the running of the GA and we will now proceed to  discuss the implementation in detail.

\paragraph*{Evolution.}
In a nutshell, the GA mimics biological evolution by emulating the notion of natural selection, which is implemented by the genetic operations of crossover and mutation.
A family of test functions evolves over a number of generations under the stochastic operators of crossover, i.e the joining of two or more functions to form another one, and mutation, i.e a random change of a function.
Then only the functions that are fittest to the data survive and are allowed to produce offspring via crossover and mutation, thereby mimicking evolutionary dynamics of a biological system under an ecological stress.
This iterative process is repeated hundreds of times to ensure convergence, and with several independent random seeds as initial conditions, in order to further explore functional space.

In order to illustrate the action of the two stochastic operators, we will now provide a simple example. Consider the two functions $f_0(x)=1+x+x^2$ and $g_0(x)=\sin(x)+\cos(x)$, then the mutation operation will randomly modify the terms in these expressions, e.g. the resulting mutated functions might be $f_0(x)\rightarrow f_1(x)=1+2x+x^2$ and $g_0(x)\rightarrow g_1(x)=\sin(x^2)+\cos(x)$, where in the latter $x=x^1$ was modified to $x^2$ inside the $\sin()$, while in the former the coefficient of the last term was altered from one to two.
On the other hand, the crossover operation randomly joins the two functions to create two more, e.g. the result might be that the terms $1+2x$ from $f_1$ and $\cos(x)$ from $g_1$ were joined to $f_1(x)\rightarrow f_2(x)=1+2x+\cos(x)$, while the other terms merged to $g_1(x)\rightarrow g_2(x)= x^2+\sin(x^2)$. 

As the GA is a stochastic approach, then the chance that a group of functions will produce offspring is proportional to its fitness to the data, which in our analysis this fitness is assumed to be a $\chi^2$ statistic. As in the GA the probability to have offspring and the fitness is proportional to the likelihood, this will then create ``evolutionary" pressure that favors the best-fitting functions, thus pushing the overall fit towards the minimum in a few generations.

\paragraph*{The specifics.}
In this analysis we reconstruct the inflationary potential  $V(\phi)$ directly from the CMB data, and the procedure to its reconstruction proceeded as follows. First, our predefined grammar consisted on the following basis of elementary functions: exp, log, polynomials etc. and a set of operations $+,-,\times,\div$, see Table \ref{tab:grammars} for a complete list.

\begin{table}[!t]
\caption{The grammars used in the GA analysis. Other complicated forms are automatically produced by the mutation and crossover operations as described in the text.\label{tab:grammars}}
\begin{centering}
\begin{tabular}{cc}
 Grammar type & Functions \\ \hline
Polynomials & $c$, $x$, $1+x$ \\
Fractions & $\frac{x}{1+x}$\\
Trigonometric & $\sin(x)$, $\cos(x)$, $\tan(x)$\\
Exponentials & $e^x$, $x^x$, $(1+x)^{1+x}$ \\
Logarithms & $\log(x)$, $\log(1+x)$
\end{tabular}
\par
\end{centering}
\end{table}

We also required that all functions reconstructed by the GA are continuous and differentiable, without any singularities in the observable region probed by the CMB data, thus avoiding in this manner spurious reconstructions or any overfitting. Here we point out again that the choice of the grammar and the population size has already been tested extensively in Ref.~\cite{Bogdanos:2009ib}.
In the same manner, the seed numbers play also a big role since they are used to create the initial population of functions used later on by the GA.

Once the initial population has been constructed, the fitness of each member is evaluated by a $\chi^2$ statistic, using the Planck data directly as input. Afterwards, using a tournament selection, see Ref.~\cite{Bogdanos:2009ib} for more details, the best-fitting functions in each generation are chosen and the two stochastic operations (crossover and mutation) are used. In order to assure convergence, the GA process is then repeated thousands of times and with various random seeds, so as to properly explore the functional space. Then the final output of the code is a smooth and analytic function of the scalar field $\phi$ that describes the potential $V(\phi)$.

Normally, the error estimates of the GA reconstructed best-fit function are determined via a path-integral approach, as originally implemented in Refs.~\cite{Nesseris:2012tt,Nesseris:2013bia}.
This approach consists in finding an analytical estimate of the error of the reconstructed quantity by calculating a path integral over all possible functions around the best-fit GA that may contribute to the likelihood; whether the data points are correlated or not.
This particular error reconstruction method has also been compared against a bootstrap Monte Carlo by Ref.~\cite{Nesseris:2012tt} and was found to be in excellent agreement.
However, in our analysis we do not have the freedom to do this kind of error reconstruction as the Planck likelihood implements a Bayesian component-separation method in pixel space, which in practice prevents one from performing the path integral.
Instead, we turn to the more computationally expensive approach of considering several independent seeds.
Then, we keep the $68.3\%$ best of all the functions in the last generation (across all different seeds), after ordering them in terms of their $\chi^2$, as a proxy to the error scatter. 

Finally, a common concern in these kinds of agnostic and parameter-independent analyses with approaches similar to the GA is the possibility of overfitting, which usually occurs when the length of the GA expressions grows arbitrarily especially when combined with composition of functions. In our case, we mitigate the risk of overfitting by keeping the length of the expressions fixed to a predetermined level, albeit at a potential cost of the adaptability of the code.
We will also perform \textit{a posteriori} checks that no overfitting is present in the best-fit function.

\section{The data and methodology \label{sec:methods}}
In this section we now describe the data and the methodology used in our analysis. Specifically, we include the temperature and $E$-polarization CMB angular power spectra of the Planck legacy release 2018 \texttt{plikTTTEEE+lowl+lowE}, which also incorporates low multipole data ($\ell < 30$). We refer to this data combination as ``Planck" \cite{Planck:2018vyg,Planck:2019nip}. We also include the Planck 2018 lensing likelihood \cite{Planck:2018lbu}, constructed from measurements of the power spectrum of the lensing potential and we refer to this dataset as ``lensing". 

Both of these data sets are used in order to perform non-parametric and local reconstructions of the inflationary potential $V(\phi)$, as described in detail in what follows.
However, even though these data are adequate for simple reconstructions, they are actually binned in terms of the multipoles $\ell$ and as such, they are insufficient for fast-oscillatory feature searches, which require the full information from the spectra. Thus, for the feature searches we will also consider the full CMB data \texttt{plik\_rd12\_HM\_v22b\_TTTEEE\_bin1.clik}, that we refer to as the ``unbinned" likelihood. Finally, we refer to the actual Planck CMB likelihood used, regardless of the data combinations, as \texttt{clik}. Next, we describe our methodology, which contains several steps, and we also present a flowchart summarizing the methodology in Fig.~\ref{fig:flow}.  

First, the GA creates a set of random potentials $V_i(\phi)$, $i=1,\ldots, N$ given a particular grammar and a given physical prior.
The number of functions $N$ is equal to the size of the population (in practice we set it to $N=100$), while the grammar may contain various combinations of the elementary functions of Table~\ref{tab:grammars}, chosen in order to facilitate the exploration of various features in the potential.

Second, we use an inflationary solver that, from each of these potentials, computes the primordial power spectrum numerically and exactly. In particular, no slow-roll approximation is used for the inflationary dynamics and the primordial power spectrum does not have to be close to scale-invariance. Initial conditions do verify the slow-roll approximation for definiteness. We also stress that our reconstruction is local in the sense that the GA is only sensitive to the potential within the observable window, and therefore cannot be extrapolated until the end of inflation.
In practice for this work, we have used the primordial module of the code \texttt{CLASS} \cite{Blas:2011rf} that we shall use in the next step any way. Note however that any other inflationary solver could be used; the versatility of the numerical method presented in this work is important as it can easily enable one to go beyond the simplest scenarios by considering, e.g., multifield models of inflation.

Third, the primordial power spectrum, independently of how it is computed, is given as an initial condition to the Boltzmann code \texttt{CLASS}, which then calculates the theoretical predictions for the temperature and polarization CMB power spectra, i.e. the $C_\ell$s.

Fourth, after we have the spectra, we pass them along to \texttt{clik}, i.e. the Planck likelihood, which then provides us with the $\chi^2$ value for each particular model. Within the context of our GA analysis this $\chi^2$ plays the role of the fitness of each ``individual" in the population, and acts as an evolutionary pressure forcing the functions to improve and adapt to the data.

Finally, after we have calculated the fitness of each individual we perform the selection and the two genetic operations of crossover and mutation, as discussed in the previous Section. Then, this procedure is iterated several hundreds of times, in this work 700, until convergence is reached, thus resulting in a best-fit function. 

Our methodology is then as follows. As we want to focus on the inflationary physics, we follow Ref.~\cite{Peiris:2013opa} and we restrict our analysis to a flat \lcdm model, where all the non-inflation related parameters are fixed to their Planck 18 best-fit values.\footnote{We use the best-fit values of parameters found in the table \texttt{base\_plikHM\_TTTEEE\_lowl\_lowE\_lensing} located at this \href{ https://wiki.cosmos.esa.int/planck-legacy-archive/images/9/9c/Baseline\_params\_table\_2018\_95pc.pdf}{URL} and in the .minimum files of the Markov chain Monte Carlo (MCMC) chains COM\_CosmoParams\_fullGrid\_R3.01.zip found in the official Planck Legacy Archive \href{https://pla.esac.esa.int/\#cosmology}{https://pla.esac.esa.int/\#cosmology}.}
In the following, we therefore only vary the inflationary potentials.
Although one might expect some small changes in the results due to the correlation of the other \lcdm parameters, we make this choice as otherwise the analysis becomes intractable.

Another, albeit more practical reason for this is that for every potential the GA creates it would be appropriate to determine the best-fit \lcdm non-inflationary parameters (either via minimization or an MCMC analysis).
However, it is clear that the analysis would become very quickly unfeasible  since in the best-case scenario we need $\mathcal{O}(10^3)$ evaluations of the likelihood to determine the best-fit \lcdm and nuisance parameters, then for $N=100$ functions in every generation (at a minimum) and for $\mathcal{O}(10^3)$ generations in total, we would require $\mathcal{O}(10^8)$ calls to \texttt{CLASS} and \texttt{clik} respectively, which is intractable with current computers. 

We also mention some caveats about our our implementation of the GA.
First, in the \texttt{CLASS} primordial module all energy scales are given in terms of the Planck mass $m_{\mathrm{pl}}$ and not of the more conventional reduced Planck mass $\Mp$.
This means that extra powers of $\sqrt{8\pi}\simeq 5.013$ appear in the equations and that we have to take them into account when translating the results back from the code and interpreting theoretically the results.
Furthermore, the primordial module of \texttt{CLASS} is sometimes unstable and frequently crashes if the power spectrum deviates a lot from the nearly flat spectrum (in which case the derived parameters for the amplitude and tilt would be vastly different).
The same also applies to the \texttt{clik} code, which frequently crashes if the model is many sigmas away from the Planck best-fit.
We have dealt with both of these issues by including error-handling code that catches the errors and penalizes these unphysical models with an extremely large $\chi^2$.

\begin{figure*}[!t]
    \centering
    \includegraphics[width=0.495\textwidth]{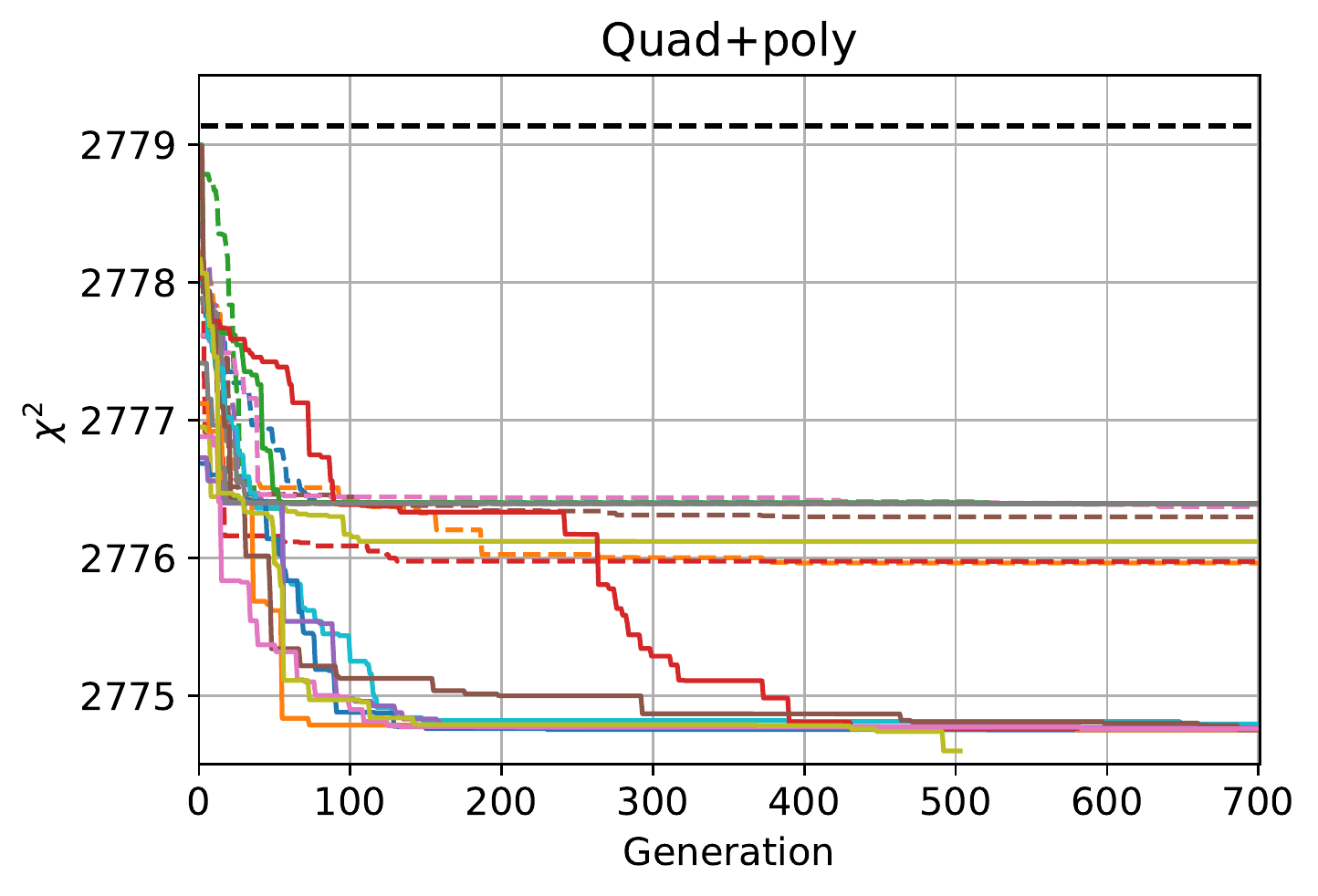}
    \includegraphics[width=0.495\textwidth]{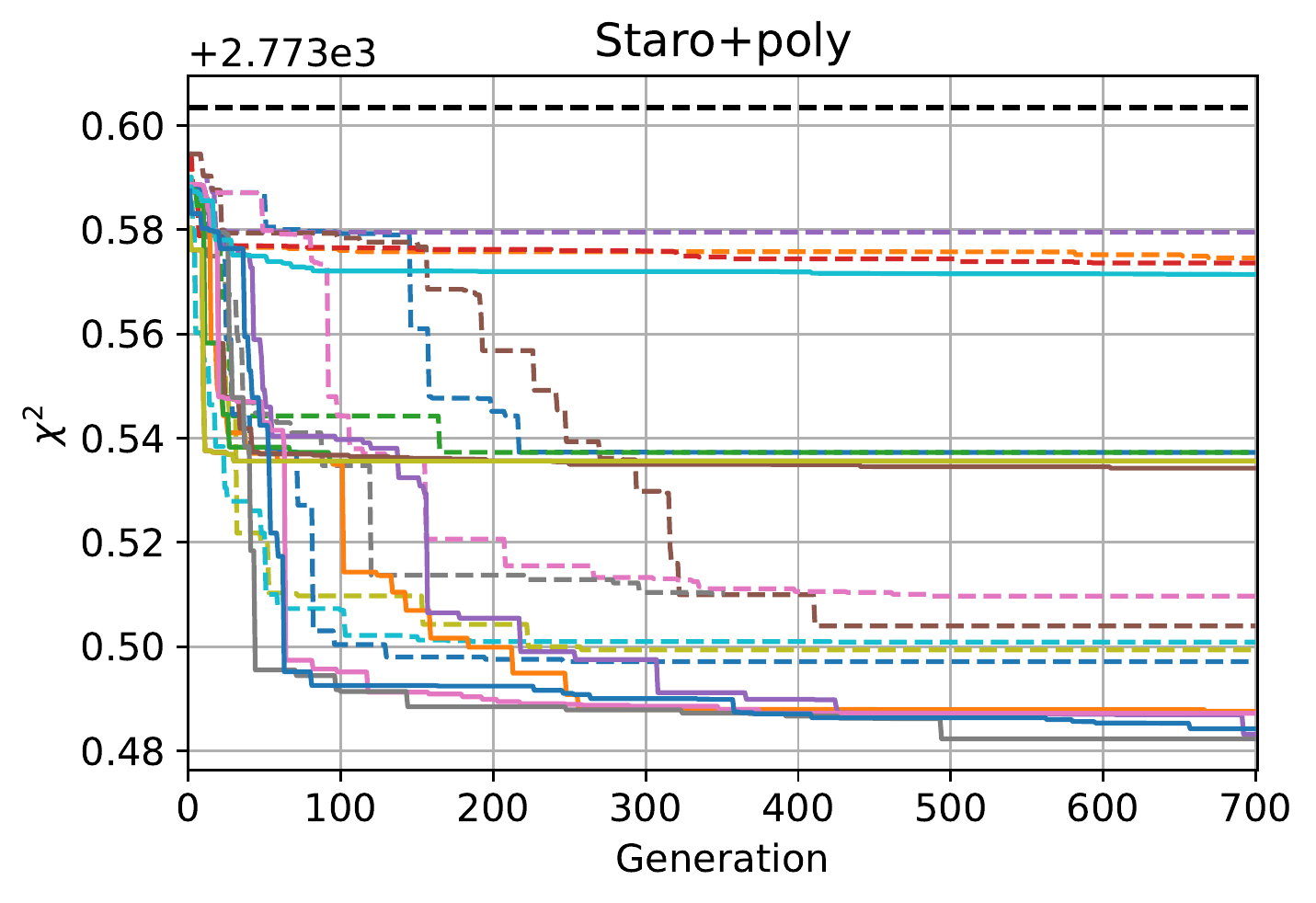}
    \caption{Typical runs of the GA showing the minimum $\chi^2$ in the population as a function of the generation.
    Individual colored lines correspond to independent runs with different random seeds for the initial conditions.
    Dashed lines correspond to the run with $\alpha=0.4$ and solid lines to the one with $\alpha=1$.
    The black dashed line represents the $\chi^2$ of the prior, shown for comparison.
    \textbf{Left panel:} Quadratic inflation model as a prior, and polynomials chosen as a grammar. The GA easily improves upon the prior, although some independent runs perform better than other ones. Larger steps in functional space, for $\alpha=1$, help the GA to find a better best-fit.
    \textbf{Right panel:} Starobinsky inflation model as prior, and polynomials chosen as a grammar.
    The GA hardly improves on the prior which is already well-fitted to the data, as can be seen from the modest value of the $\Delta\chi^2$.
    Changing $\alpha$ does not help, showing the robustness of the prior with respect to the data.
    In both panels and for all cases, the GA converges to a minimum.}
    \label{fig:chi2s}
\end{figure*}

\begin{figure*}[!t]
    \centering
    \includegraphics[width=0.475\textwidth]{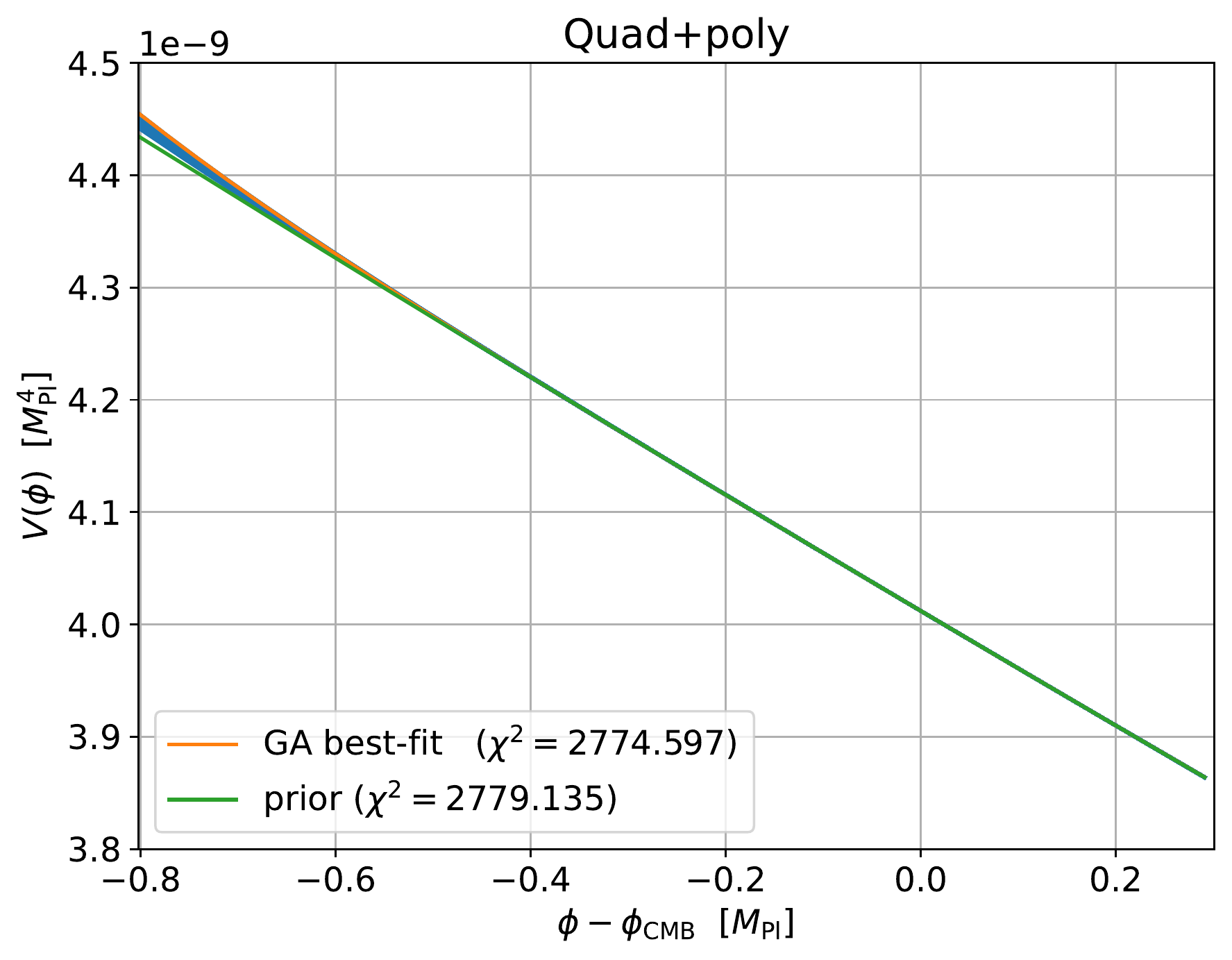}
    \includegraphics[width=0.495\textwidth]{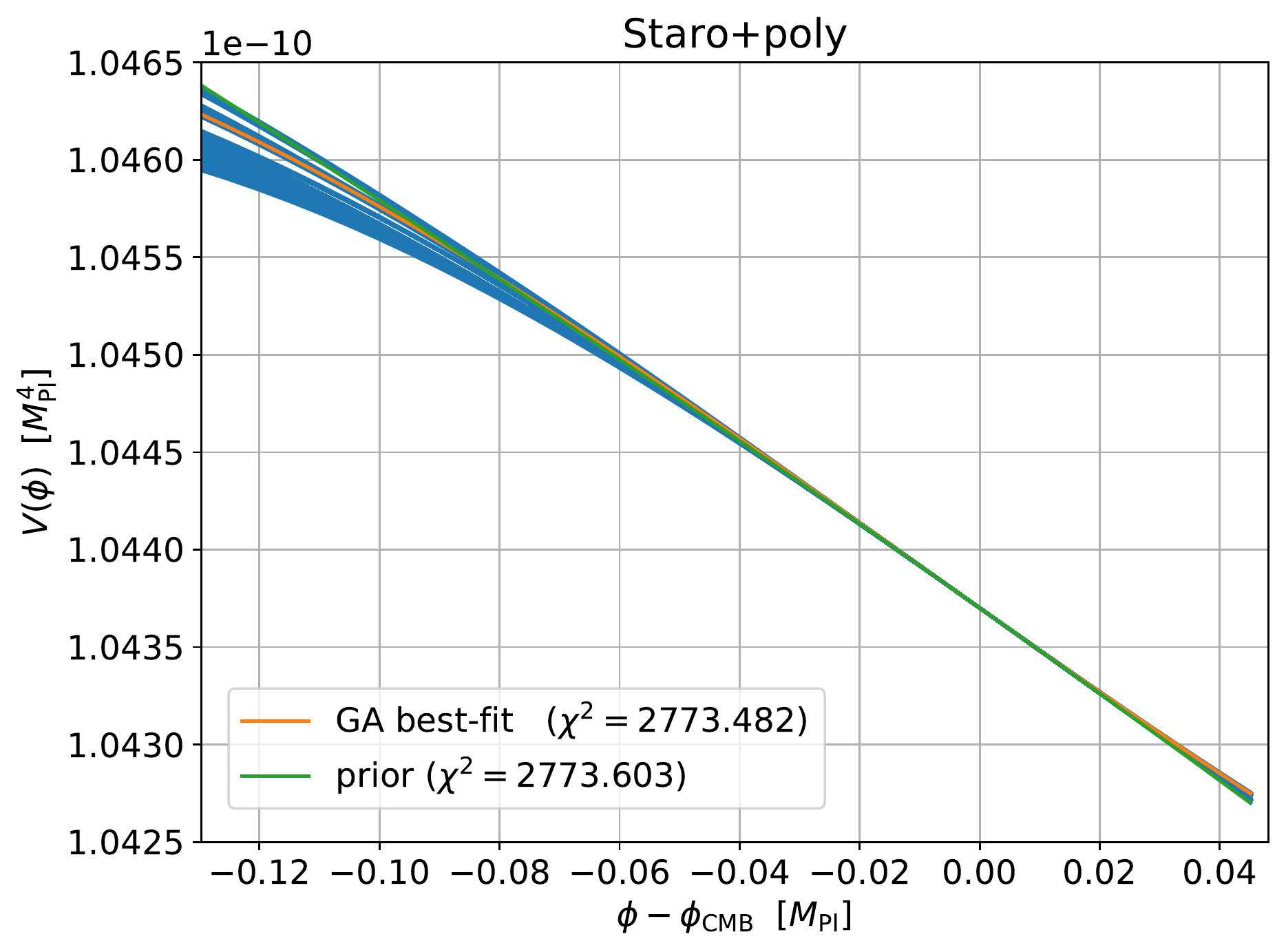}
    \includegraphics[width=0.48\textwidth]{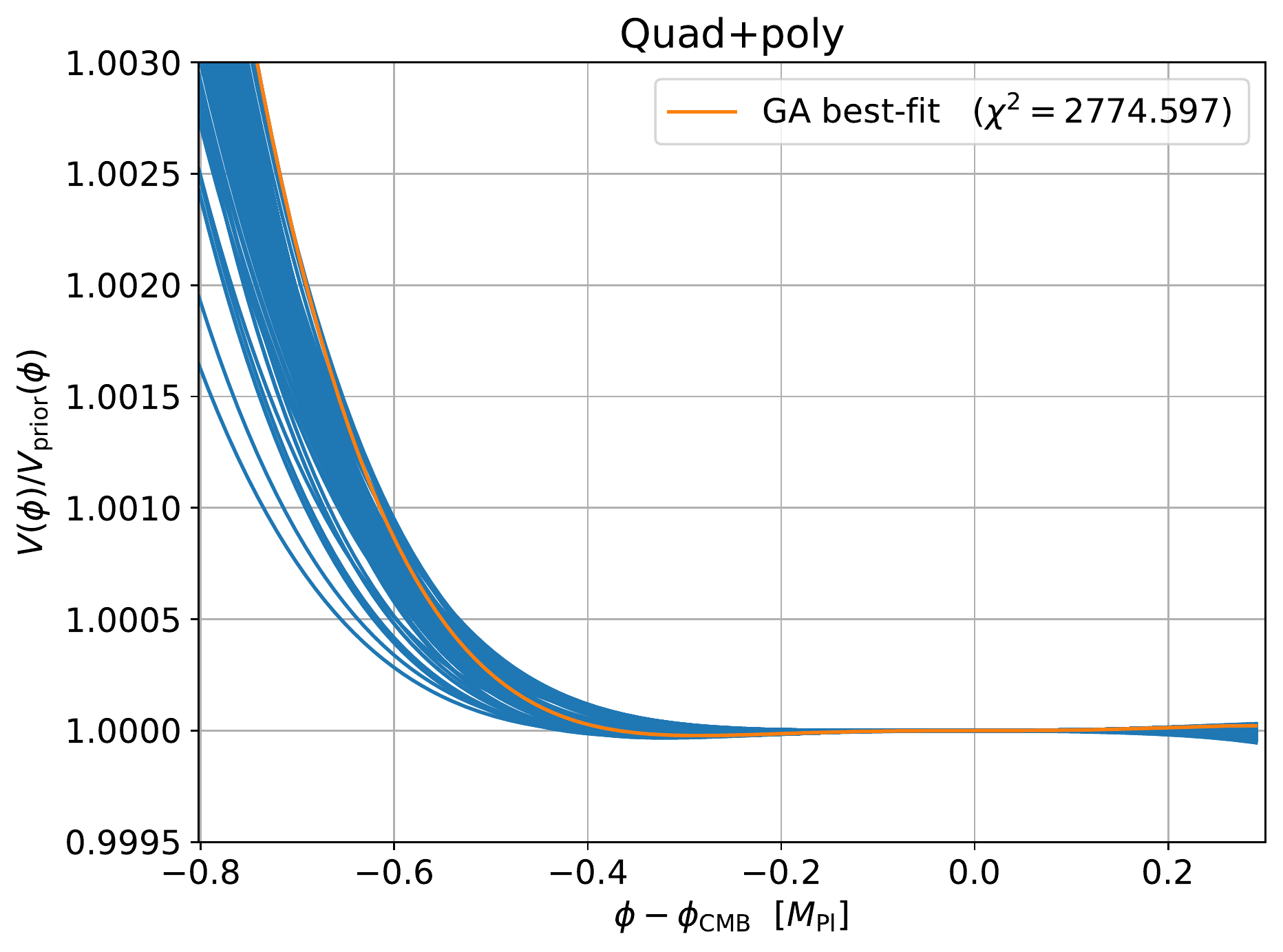}
    \includegraphics[width=0.495\textwidth]{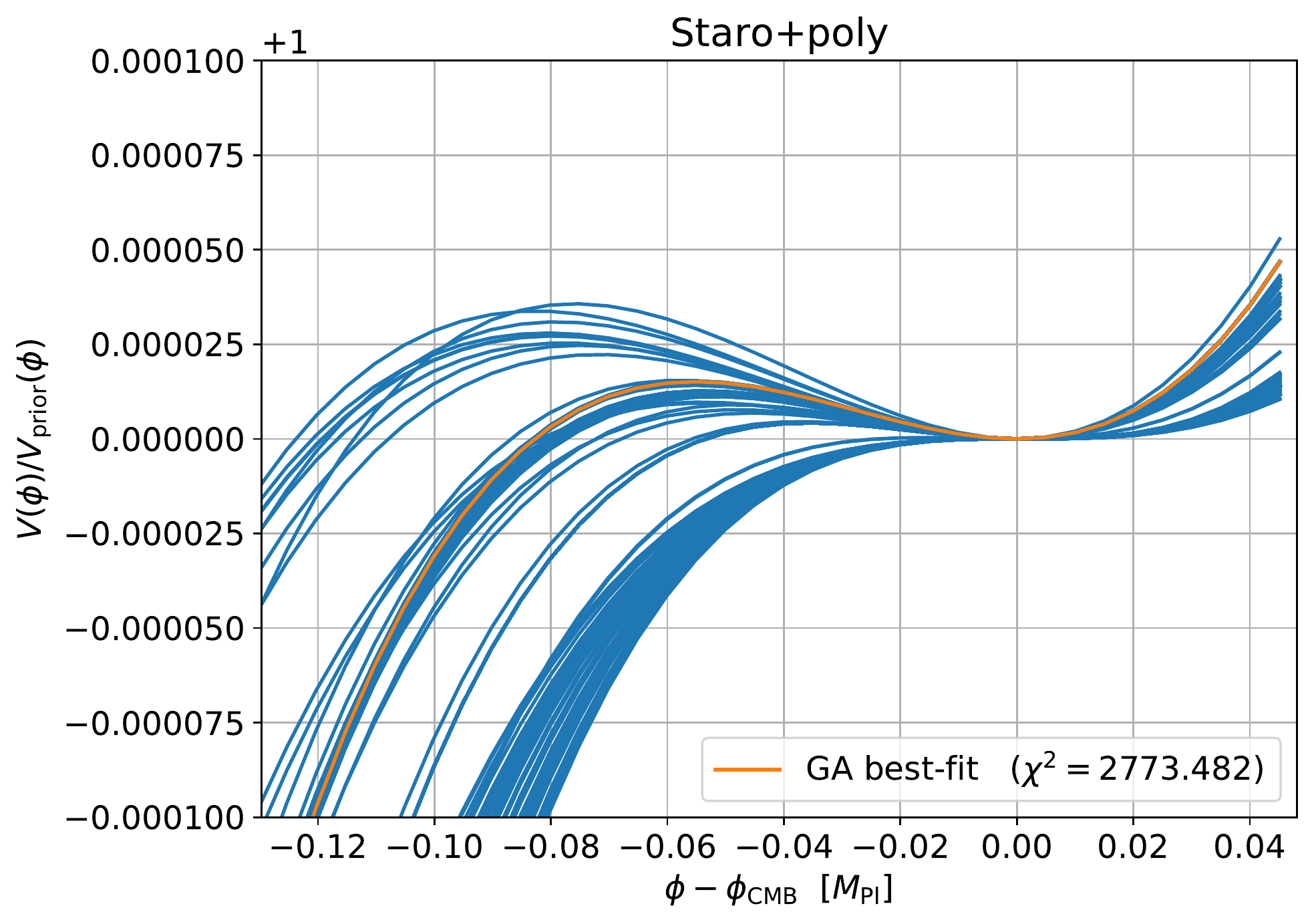}
    \caption{Inflationary potentials reconstructed by the GA at the last generation (blue lines), including the best-fit function (orange line), and with the prior shown for comparison (green line).
    The upper panels show the total potential constrained by data, the lower panels show the same potential but rescaled by the prior and zoomed around unity.
    \textbf{Left panels:} Quadratic inflation model as a prior, polynomials chosen as a grammar, and $\alpha=1$. 
    \textbf{Right panels:} Starobinsky inflation model as prior, polynomials chosen as a grammar, and $\alpha=1$.
    As described in the text, the scatter of the potentials (blue lines) around the best-fit is a proxy for the error estimate of our reconstructions: here we only show the $68.3\%$ best of all the functions in the last generation (across all seeds), after ordering them in terms of their $\chi^2$.}
    \label{fig:bestfits-slow-roll}
\end{figure*}

\section{Results \label{sec:results}}
In this section, we apply the methodology of the previous section to the construction of inflationary potentials of the form:
\begin{equation}
    V(\phi)=V_\mathrm{prior}(\phi)\left[1+\alpha\,\mathrm{GA}(\varphi)\right]\,,
\end{equation}
where we have set  $\varphi\equiv(\phi-\phi_\mathrm{CMB})/\Mp$ with $\phi_\mathrm{CMB}$ being the value of the scalar field when the pivot scale in the CMB has exited the horizon.
Indeed, the GA is reconstructing potentials only \textit{locally} around this value of $\phi$ and cannot be extrapolated further. 

Depending on the direction we want to explore, we may choose different physical priors for the potential, different data sets of Planck (binned or unbinned), and different grammars for the GA.
The parameter $\alpha$ represents the size of the steps around the prior that the GA can perform in order to explore the functional space. 
In what follows, we will perform runs with  different values for $\alpha$, namely $\alpha=0.4$ (in order to keep the ``distance" from the prior modest) and $\alpha=1$ (in order to explore more thoroughly the functional space).
See Table~\ref{tab:best_GA} for a summary of the results we obtain for different data sets, priors, values of $\alpha$ and grammar.

\subsection{Slow-roll potentials}
We first investigate how the GA can be used to reconstruct slow-roll potentials, i.e. monotonic functions $V(\phi)$ featuring small derivatives at any order, in the observable region surrounding $\phi_\mathrm{CMB}$, the value of the scalar field when the pivot scale of the CMB, $k_\star=0.05\,\mathrm{Mpc}^{-1}$, exits the Hubble radius. 
For this purpose, we will consider two kinds of physical priors for the GA, corresponding to different initial conditions for the potential:
\begin{enumerate}
    \item Quadratic inflation with $V_\mathrm{quad}(\phi)=m^2\phi^2/2$, the simplest vanilla model of slow-roll inflation. We will use the current best-fit to the Planck data, corresponding to $m=5.736\times 10^{-6}\,\Mp$, with a  $\chi^2_\mathrm{quad}=2779.135$. In practice, only the observable window around $\phi_\mathrm{CMB}= -15.616\,\Mp$ is explored by data.
    \item Starobinsky inflation with $V_\mathrm{Staro}(\phi)=V_0\,\left[1-\mathrm{exp}\left({\sqrt{\frac{2}{3}}\frac{\phi}{\Mp}}\right)\right]^2$, whose best-fit corresponds to $V_0=1.071\times 10^{-10}\Mp^4$ and for which $\phi$ around $\phi_\mathrm{CMB}=-5.356\,\Mp$ is explored by data, with  $\chi^2_\mathrm{Staro}=2773.603$.\\~~\\   
    For the sake of completeness, we will also consider another small-field model of inflation with concave potential, the so-called hilltop model with $V_\mathrm{hilltop}=V_0\left[1-\phi^2/\Lambda^2\right]$, whose best-fit corresponds to   $\Lambda=10.976\,\Mp$ and $V_0=1.084\times 10^{-10}\,\Mp^4$, and for which $\phi_\mathrm{CMB}=1.241\,\Mp$, with $\chi^2_\mathrm{hilltop}=2773.577$.
\end{enumerate}

\paragraph*{Beating quadratic inflation.}
We start by the case corresponding to $V_\mathrm{prior}(\phi)=V_\mathrm{quad}(\phi)$, restricting the analysis to the binned data sets of Planck, and allowing only for polynomials in the grammar of the GA, in order to perform the first proof of concept of the algorithm in the simplest case.
The minimum $\chi^2$ of all functions at each generation of the GA, for each of the different  random seeds, can be seen in the left panel of Fig.~\ref{fig:chi2s}.
Clearly, the GA works as expected: by creating new functions and performing the selection by fitness, the population converges towards the minimum, that performs better than the prior when compared to the  Planck data.

The best-fit function found by the GA is better for $\alpha=1$ than for $\alpha=0.4$, and corresponds to an improvement of $\Delta \chi^2=\chi^2_\mathrm{GA,min}-\chi^2_\mathrm{prior}=-4.538$.
It is plotted in the left panels of Fig.~\ref{fig:bestfits-slow-roll}, and is given analytically by the GA:
\bea
\label{eq:besfit-quad-prior}
V_\mathrm{GA,min}/V_\mathrm{quad}&= & \, 1+\,2.442\times 10^{-3}\, \varphi ^3-1.824\times 10^{-2}\, \varphi ^5\nn \\
&-& 6.238\times 10^{-4}\,\varphi^6.
\eea 
By inspecting the theoretical predictions for the CMB power spectra of both the best-fit and the quadratic inflation potential prior and comparing with the Planck 18 data points, we can see that the improvement in the $\chi^2$ mainly comes from a better fit of the GA to the low-$\ell$ temperature $C_\ell^{TT}$, and to a much lesser extent also from the  low-$\ell$ temperature-polarization cross-correlation $C_\ell^{TE}$.

Note that the potential found by the GA is still not quite as good as the best single-field slow-roll models, like Starobinsky inflation.
In order to find a best-fit with an even better $\chi^2$ with the GA and starting from the quadratic inflation prior, one should carry an extensive study by changing several parameters of the GA. For example, one could increase the population size from $N=100$, in order to allow for more genetic variety and better combinations.
One could also change the mutation rate as well as other internal parameters of the GA.
As this work is a proof of concept, we leave for future work the full investigation of the functional space explored by the GA, done by tweaking its various parameters, e.g. having  bigger population sizes and different mutation and crossover rates, in order to fully optimize the best-fit potential.

\paragraph*{Robustness of Starobinsky inflation.}
We now change prior and consider the Starobinsky inflation model as an initial condition for the algorithm.
In the right panel of Fig.~\ref{fig:chi2s}, we show the minimum $\chi^2$ at each generation for each random seed, when the grammar is chosen to be polynomials.
As is clear, this time the GA hardly beats the prior, as for example the best-fit candidate has a $\Delta\chi^2=-0.106$ for $\alpha=0.4$.
This result is consistent with the fact that this model is known for predicting $n_s$ precisely in the observationally allowed range and with a very small value of $r$, which is also consistent with the absence of detection of primordial $B$-modes so far.

In Table~\ref{tab:best_GA}, we show the best $\Delta\chi^2$ when changing the grammar from polynomials to trigonometric functions (with order $\Mp$ wavelengths in $\phi$-space, which we call ``slow" trigonometric functions, as we want to remain within slow-roll potentials in this section), and when allowing for both polynomials and trigonometric functions.
We have also changed to $\alpha=1$ while keeping the polynomial grammar, in order to allow the GA to do larger steps in functional space.
In all cases, we find either a marginal improvement upon the prior ($\Delta\chi^2 \simeq -0.1$ at best), or no improvement at all ($\Delta\chi^2>0$), which confirms the robustness of Starobinsky inflation as an excellent single-field slow-roll model.
Actually, we stress that this is a feature shared with other small-field models of inflation: in Table~\ref{tab:best_GA} we also show that the best $\Delta\chi^2$ when using the hilltop model as a prior are again only marginally deviating from zero.

\subsection{Beyond slow-roll: oscillatory features in the potential}
We now perform a model-independent search for oscillatory features in the inflationary potential.
In order to do this, we now allow for trigonometric functions with small wavelengths in $\phi$-space (of order $10^{-4}$\,--\,$10^{-2} \Mp$, which we call ``fast" trigonometric functions) in the grammar.
\footnote{
In order to sample correctly the expected resulting oscillations in the power spectrum, we also increased the sampling rate of the primordial module from the default value to 100 points per decade.
In practice we set k\_per\_decade\_primordial = 100 in the .ini file of \texttt{CLASS}.
We also checked that increasing this number any further does not affect our results.
}

Note that these oscillations are precisely the ones encountered in axion monodromy inflation with natural values of the axion decay constant $f \ll \Mp$ (see, e.g., the review~\cite{Pajer:2013fsa}), but that crucially we do not force the GA to focus on this particular model.
Oscillatory features in the potential result in resonant features with oscillations linear in $\log k$-space in the primordial scalar power spectrum, which is known for improving the fit to CMB data with respect to the slow-roll models of inflation.

\begin{figure}[!t]
    \centering
    \includegraphics[width=0.497\textwidth]{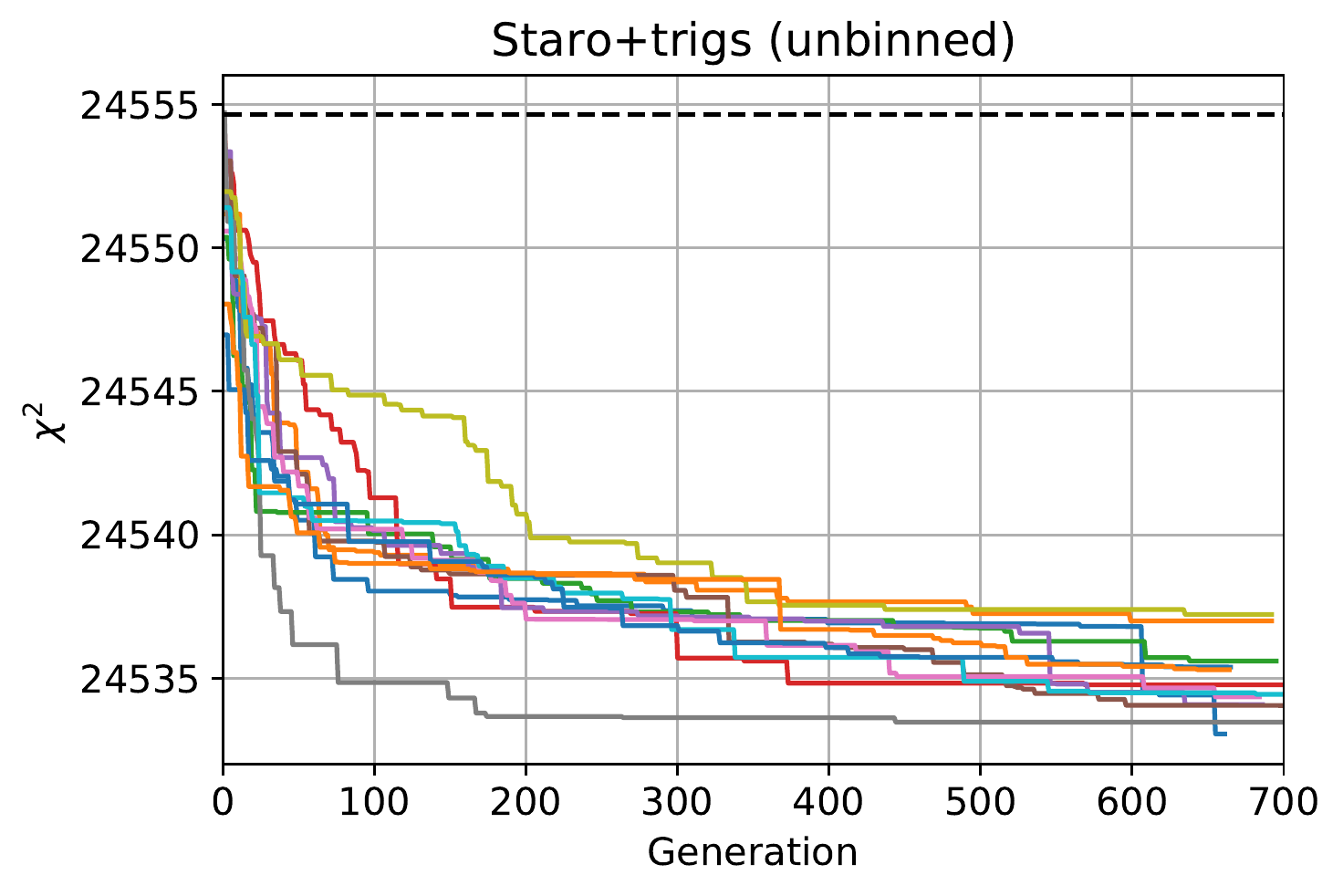}
    \caption{The evolution of the population fitness as a function of the generations for the Starobinsky prior with fast trigonometric grammar and the unbinned CMB data, and for $\alpha=1$.
    Individual colored lines correspond to independent runs with different random seeds for the initial conditions, and the black dashed line represents the $\chi^2$ of the prior, shown for comparison.}
    \label{fig:chi2-unbinned}
\end{figure}

\begin{figure}[!t]
    \centering
    \includegraphics[width=0.497\textwidth]{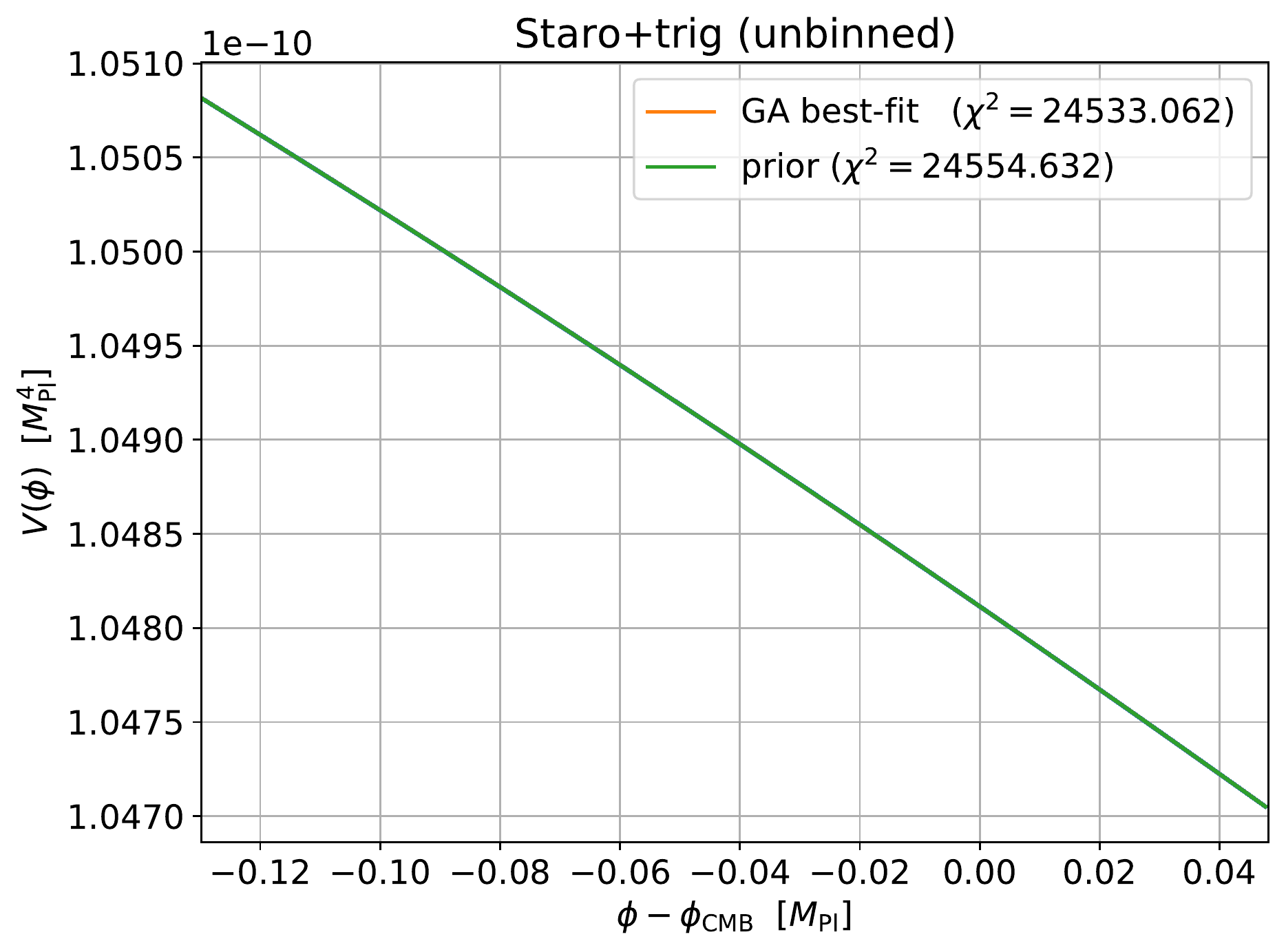}
    \includegraphics[width=0.497\textwidth]{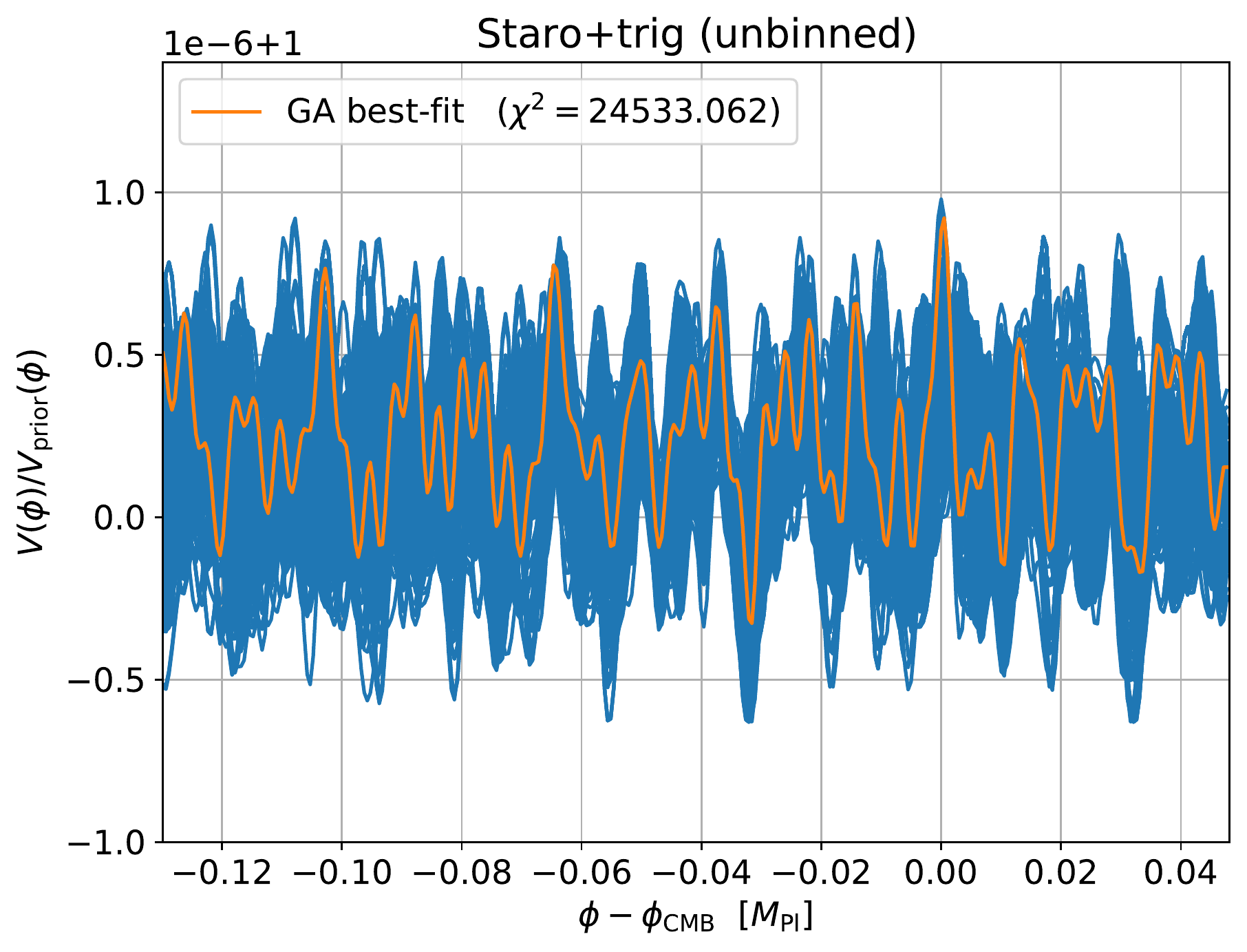}
    \caption{As in Fig.~\ref{fig:bestfits-slow-roll}, but for the Planck unbinned data using the Starobinsky inflation model as a prior, a fast trigonometric grammar and $\alpha=1$. \textbf{Upper panel:} The total potential, as constrained by the data, deviations from the prior being invisible by eye. \textbf{Lower panel:} The same potential, but rescaled by the prior and zoomed around unity.
    We recall that blue lines correspond to the $68.3\%$ best functions in the last generation across all seeds and therefore represent the error scatter. The structure of this error scatter unveils a clear pattern for these fast oscillatory functions.
    }
    \label{fig:bestfit-oscillatory}
\end{figure}

\begin{table*}[!t]
\caption{Here we present the best-fit results for the various cases considered in the analysis. In particular, we show the difference in $\chi^2$ with respect to the prior, $\Delta \chi^2=\chi^2_\mathrm{GA,min}-\chi^2_\mathrm{prior}$, thus a negative $\Delta\chi^2$ means the GA has found a better best-fit.
\label{tab:best_GA}}
\begin{centering}
\begin{tabular}{|c|c|c|c|c|c|c|}
  \hline
  Data & Prior & $\chi^2_\mathrm{prior}$ & $\alpha$ & Grammar & $\chi^2_\mathrm{GA,min}$ & $\Delta\chi^2$ \\
  \hline
  \,Binned\, & \,Quadratic\, & \,$2779.135$\, & \,$0.4$\, & \,polynomials\, & \,$2775.962$\, & \,$-3.173$\, \\
  \,Binned\, & \,Quadratic\, & \,$2779.135$\, & \,$1.0$\, & \,polynomials\, & \,$2774.597$\, & \,$-4.538$\, \\
  \,Binned\, & \,Starobinsky\, & \,$2773.603$\, & \,$0.4$\, & \,polynomials\, & \,$2773.497$\, & \,$-0.106$\, \\
  \,Binned\, & \,Starobinsky\, & \,$2773.603$\, & \,$0.4$\, & \,trigonometric (slow)\, & \,$2773.609$\, & \,$0.006$\, \\
  \,Binned\, & \,Starobinsky\, & \,$2773.603$\, & \,$0.4$\, & \,polynomials + trigonometric (slow)\, & \,$2773.491$\, & \,$-0.112$\, \\
  \,Binned\, & \,Starobinsky\, & \,$2773.603$\, & \,$1.0$\, & \,polynomials\, & \,$2773.482$\, & \,$-0.121$\, \\
  \,Binned\, & \,Hilltop\, & \,$2773.577$\, & \,$0.4$\, & \,polynomial\, & \,$2773.567$\, & \,$-0.010$\,\\
  \,Binned\, & \,Hilltop\, & \,$2773.577$\, & \,$0.4$\, & \,trigonometric (slow)\, & \,$2773.679$\,  & \,$0.102$\,  \\
  \,Unbinned\, & \,Starobinsky\, & \,$24554.632$\, & \,$1.0$\, & \,trigonometric (fast) \, & \,$24533.062$\, & \,$-21.570$\, \\
  \hline
\end{tabular}
\end{centering}
\end{table*}

\begin{figure*}[!t]
    \centering
    \includegraphics[width=0.47\textwidth]{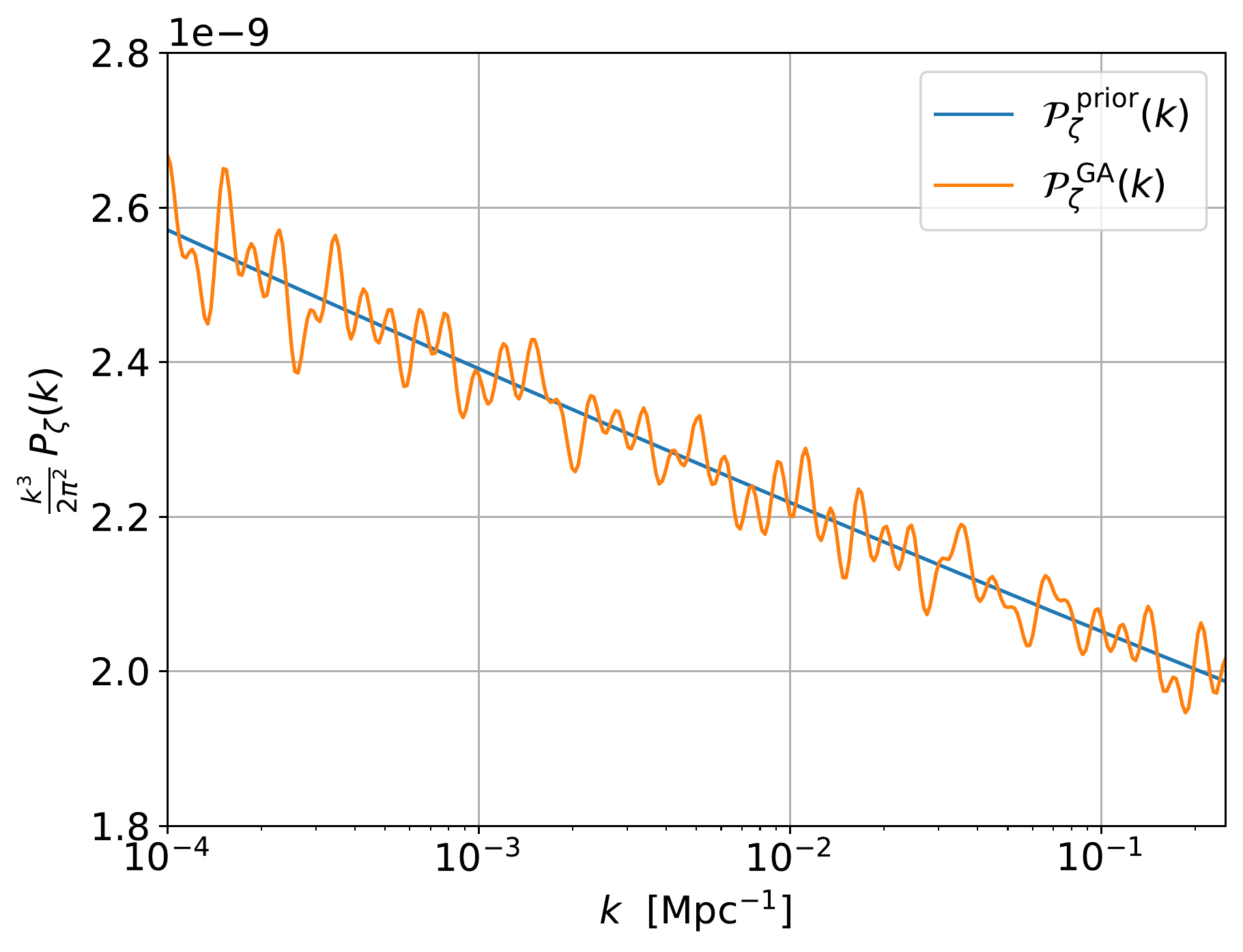}
    \includegraphics[width=0.49\textwidth]{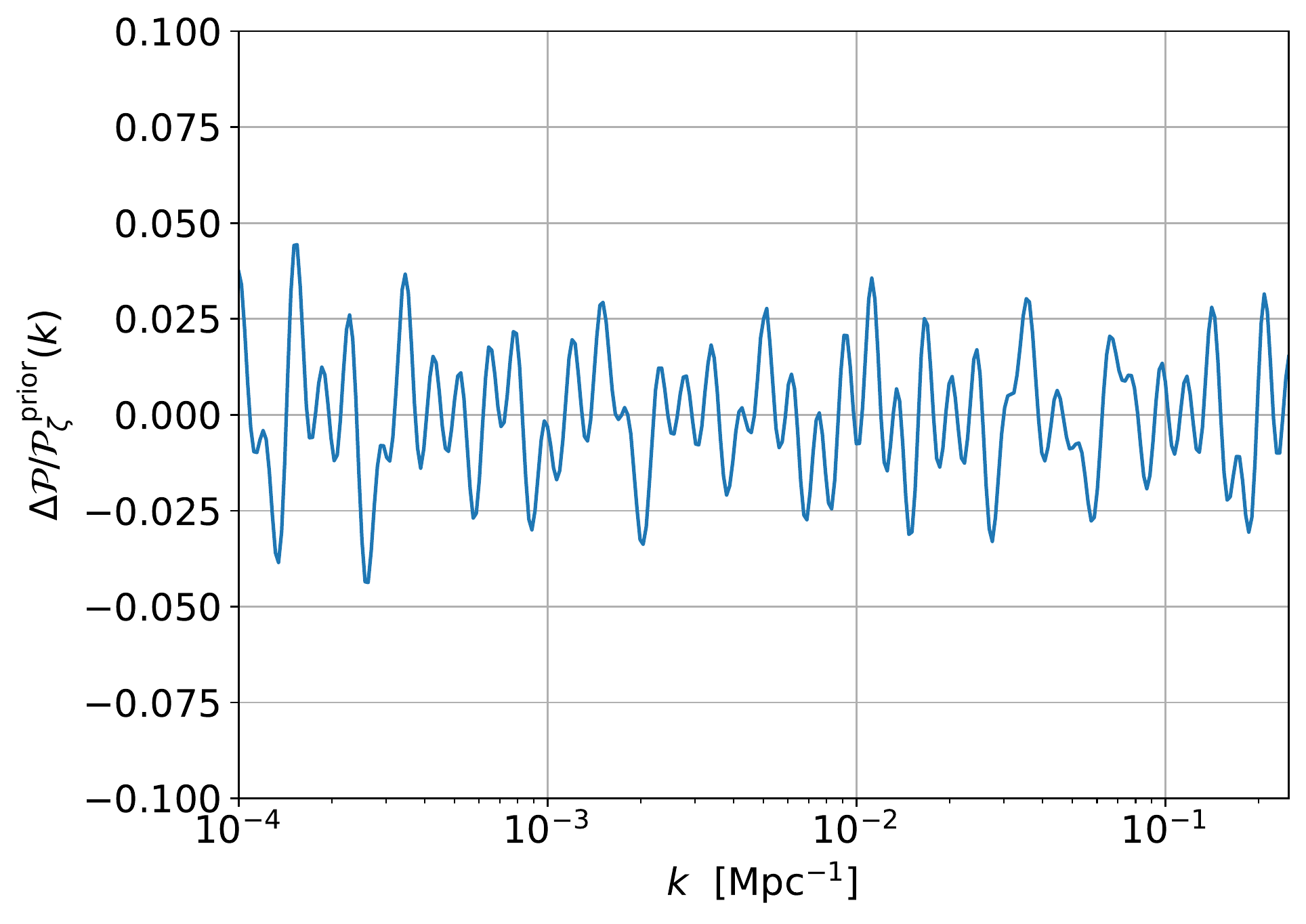}
    \caption{\textbf{Left panel:} The dimensionless primordial scalar power spectrum $[k^3/(2\pi^2)] P_\zeta(k)$ as a function of the wavenumber $k$ for the best-fit candidate found by the GA with the Starobinsky inflation model as a prior, fast trigonometric grammar, $\alpha=1$ and using Planck unbinned data (orange line). As can be seen, the oscillations in $\phi$-space from the potential are induced on the power spectrum in $\log k$-space.
    The power spectrum from the prior (blue line) is shown for comparison.
    \textbf{Right panel:} The relative variation of the best-fit power spectrum with respect to the prior, given by Eq.~\eqref{eq:DP}.}
    \label{fig:pk}
\end{figure*}

We show in Fig.~\ref{fig:chi2-unbinned} the evolution of the population fitness to the unbinned Planck data as a function of the generation, with the Starobinsky inflation model as a prior for the GA.
Strikingly, the GA now easily beats one of the best slow-roll single-field models of inflation by including oscillatory features in the potential, as can be seen from the sudden and sizable decrease of the $\chi^2$, with the best-fit final potential corresponding to $\Delta\chi^2 = -21.570$.
The best-fit function given analytically by the GA reads
\bea
    \label{eq:besfit-unbinned-staro-prior}
    V_\mathrm{GA,min}/V_\mathrm{staro} = && 1+1.235\times 10^{-7}\,\sin(1631.340\,\varphi) \nn \\
    &+& 3.961\times 10^{-7}\,\cos(245.121\,\varphi)^4 \nn \\
    &+& 2.272\times 10^{-7}\,\cos(298.822\,\varphi)^7 \nn \\
    &+& 2.598\times 10^{-7}\,\cos(430.520\,\varphi)^4\,,
\eea
and is plotted together with the rest of the population at the last generation in Fig.~\ref{fig:bestfit-oscillatory}.
To our knowledge, this is the best improvement upon slow-roll inflation when reconstructing the inflationary potential and using latest Planck unbinned data (in different contexts, see other best-fit candidates with $\Delta\chi^2 \sim -20$: Refs.~\cite{Braglia:2021rej,Braglia:2021sun} in a model-based approach and Ref.~\cite{Canas-Herrera:2020mme} for a reconstruction of the inflaton's speed of sound). 

\begin{figure}[!t]
    \centering
    \includegraphics[width=0.485\textwidth]{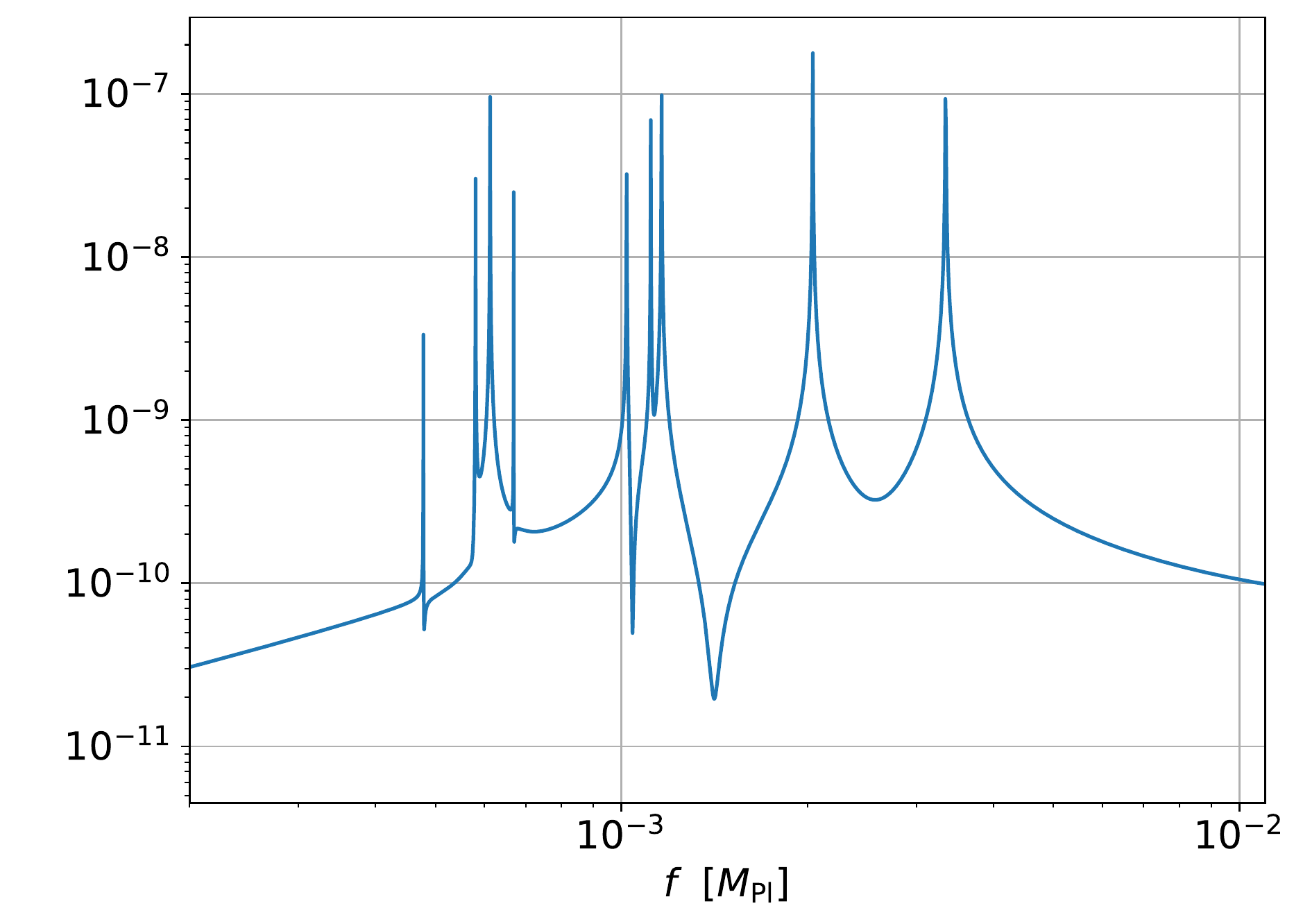}
    \includegraphics[width=0.485\textwidth]{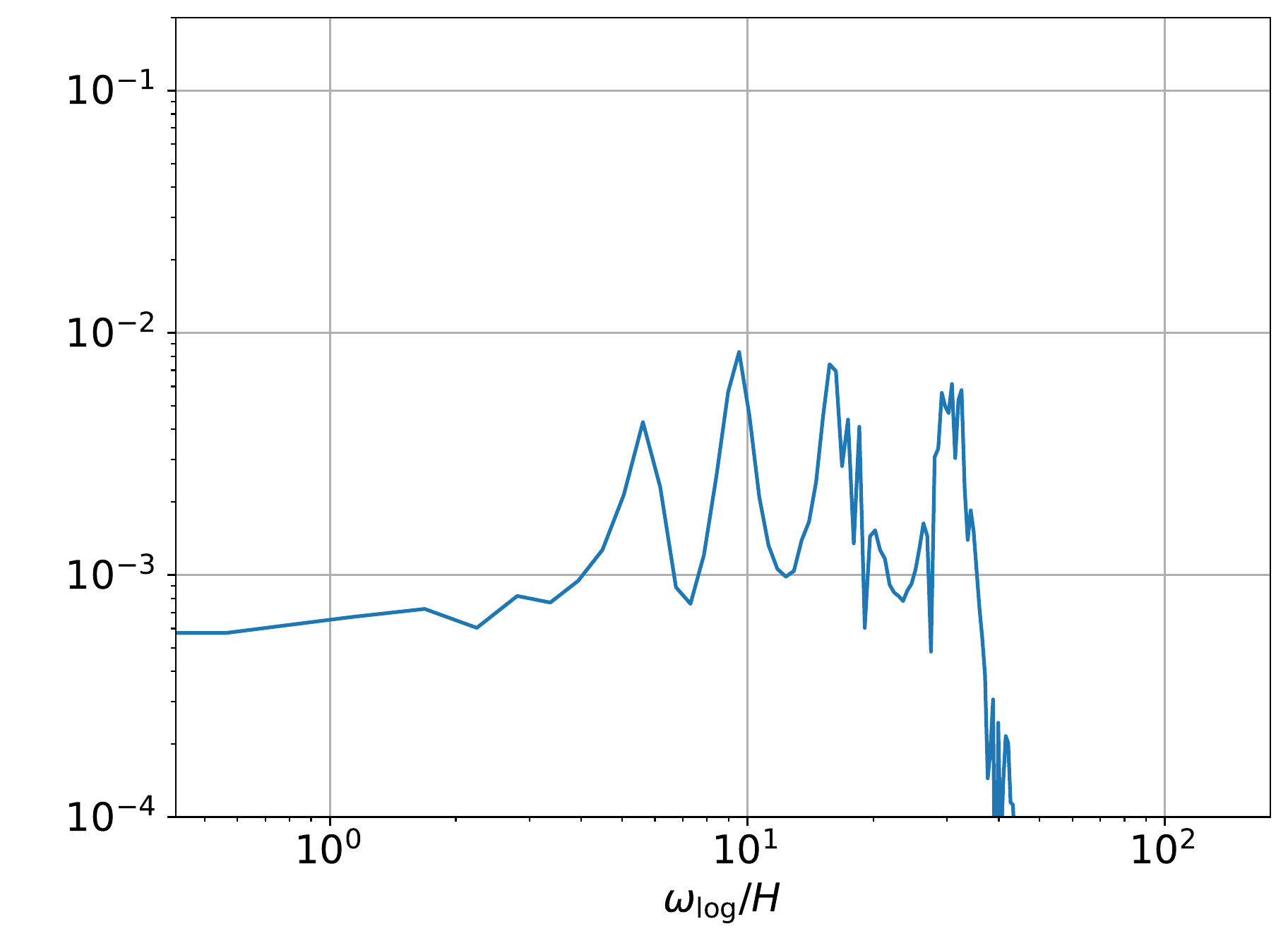}
    \caption{\textbf{Upper panel:} Fourier transform of the best-fit potential giving the multimodal distribution of frequencies in field space.
    Here in the abscissa we have represented the inverse of these frequencies, i.e. the wavelengths in $\Mp$ units, which could be interpreted as axion decay constants $f$. 
    \textbf{Lower panel:} Fourier transform of the primordial scalar power spectrum resulting from the best-fit potential, giving the distribution of frequencies in $\log k$-space, often denoted as $\omega_{\log}/H$ in the literature.}
    \label{fig:bestfit-oscillatory-fourier-transform}
\end{figure}

In Fig.~\ref{fig:pk}, we show the primordial scalar power spectrum resulting from the best-fit function given by Eq.~\eqref{eq:besfit-unbinned-staro-prior}.
The upper panel displays the power spectrum itself, while the lower panel contains the relative variation with respect to the prior, i.e. the quantity 
\be
\frac{\Delta \mathcal{P}}{\mathcal{P}_\zeta^\mathrm{prior}}(k)\equiv \frac{\mathcal{P}_\zeta^\mathrm{GA}(k)-\mathcal{P}_\zeta^\mathrm{prior}(k)}{\mathcal{P}_\zeta^\mathrm{prior}(k)}.\label{eq:DP}
\ee 
As expected, we recover the oscillations linear in $\log k$-space of resonant features, although a striking feature of this best-fit potential is that it contains multiple frequencies.
We show this multi-modal distribution of frequencies, both in $\phi$-space and in $\log k$-space, in Fig.~\ref{fig:bestfit-oscillatory-fourier-transform}. 
For comparison, the single-harmonic best-fit to data found when exploring resonant features with the conventional Monte Carlo approach at the level of the power spectrum corresponds to a frequency $\omega_\mathrm{log}/H\sim18$ in $\log k$-space (see, e.g., Ref.~\cite{Braglia:2022ftm}).

\begin{figure}[!t]
    \centering
    \includegraphics[width=0.489\textwidth]{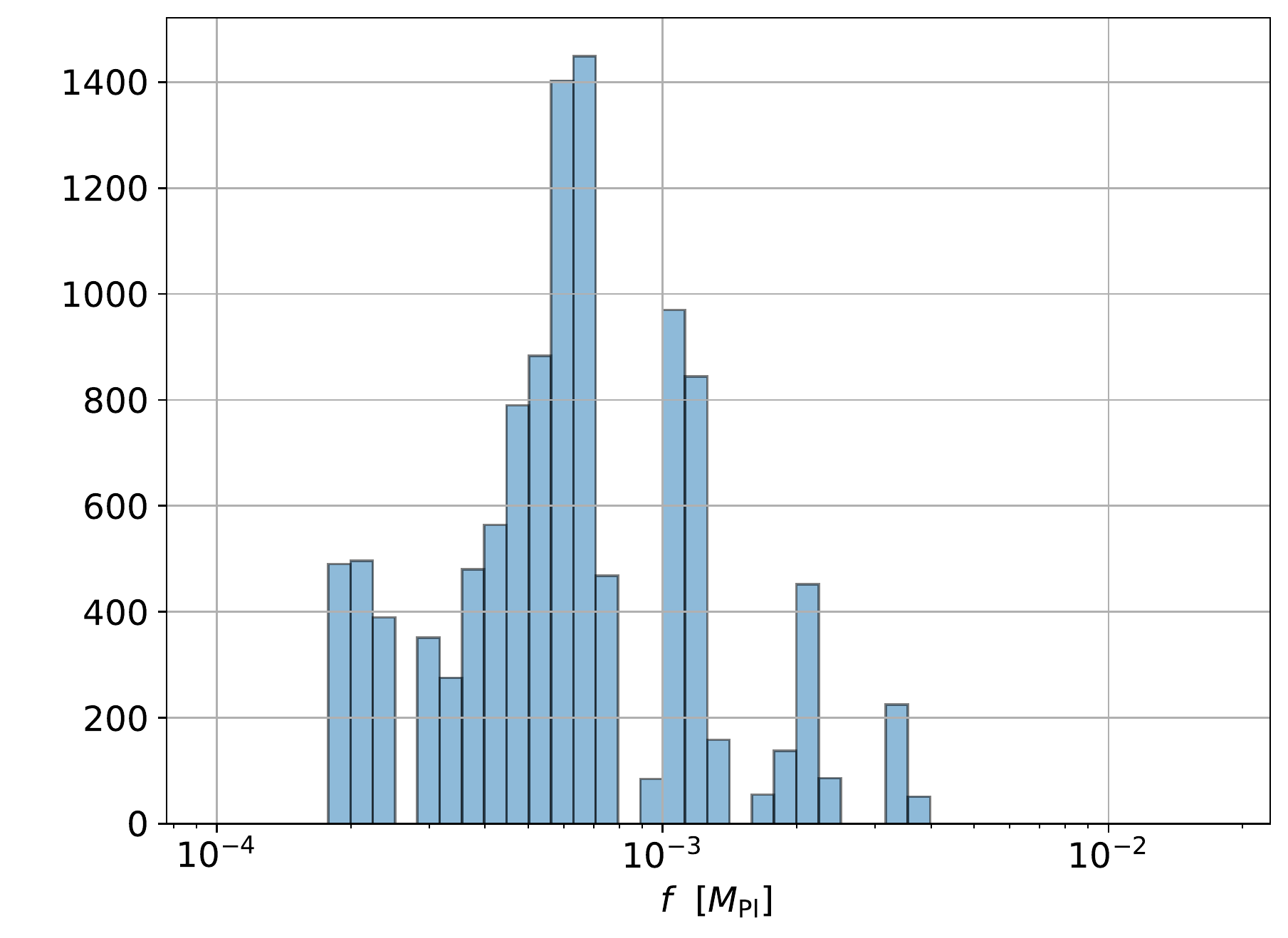}
    \caption{Distribution of the axion decay constants $f$ from the $68.3\%$ best of all functions at the last generation, across all random seeds. We used a constant binning width in log-space of $\log_{10}\left(\Delta f/\Mp \right)=0.05$.
    The structure of the histogram shows that specific values are more often encountered than other ones.
    }
    \label{fig:spectra}
\end{figure}

We now perform a statistical analysis of the frequencies found by the GA in the $68.3\%$ best-fit functions in the last generation of all seeds.
In Fig.~\ref{fig:spectra} we show a histogram of the distribution of the axion decay constants $f$ in these oscillatory functions, using a constant binning width in log-space of $\log_{10}\Delta f/\Mp=0.05$.
Clearly, some axion decay constants seem to emerge more often than others, as shown by the clustered structure around specific values.

We also recall that the risk of overfitting by the GA is
mitigated
by the inclusion of a maximal length for the inflationary potentials proposed by the algorithm, which in practice we fixed to four terms.
This choice is the result of experience gathered by using the algorithm, and consists in a sweet spot in terms of adaptability for the GA to explore functional space; in particular we have checked that reducing further the maximal length penalizes the algorithm, as we found that the best-fit to the data is then worse.

\begin{figure*}[!tp]
    \centering
    \includegraphics[width=0.82\textwidth]{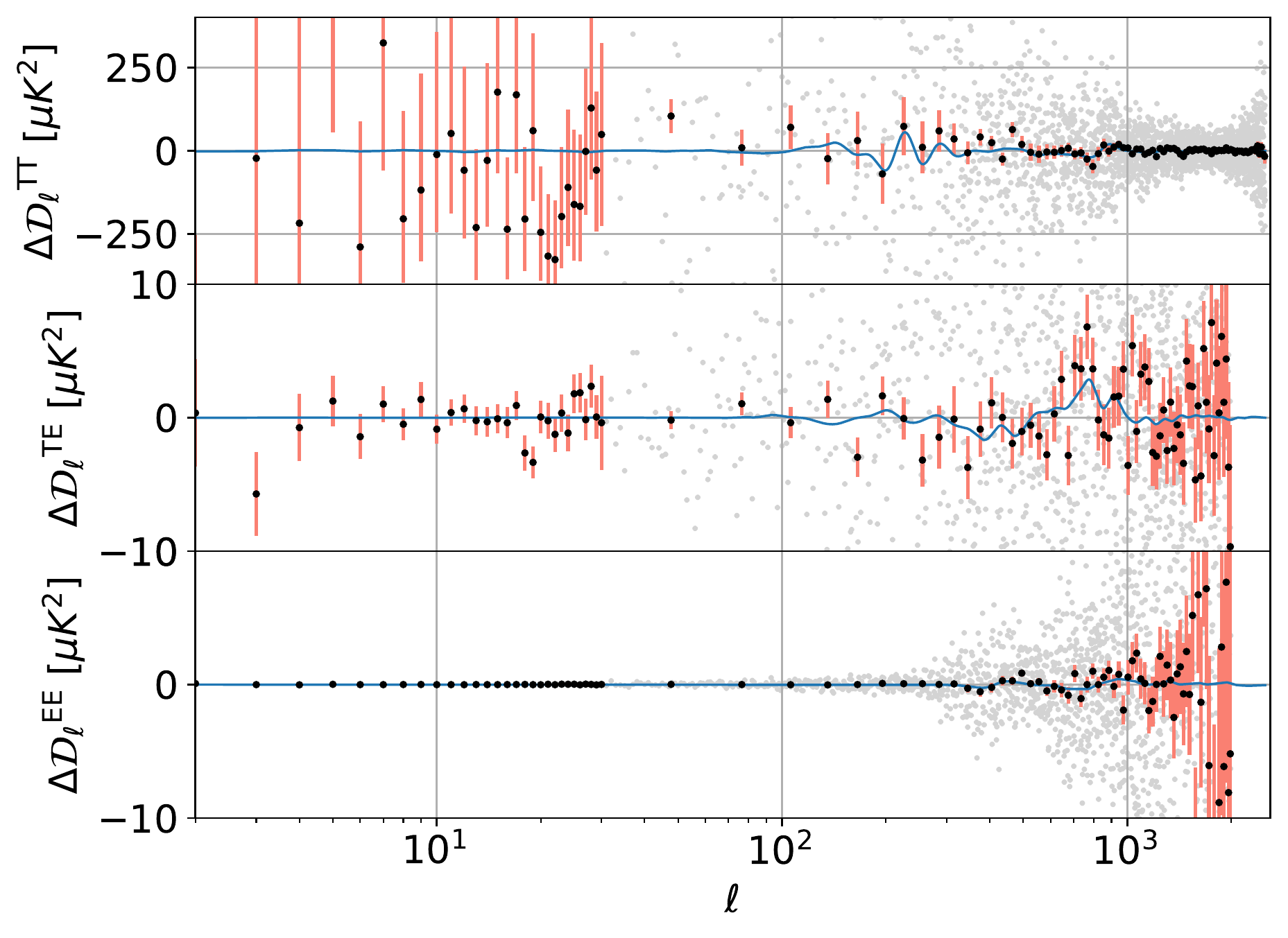}
    \hspace*{0.6cm}\includegraphics[width=0.85\textwidth]{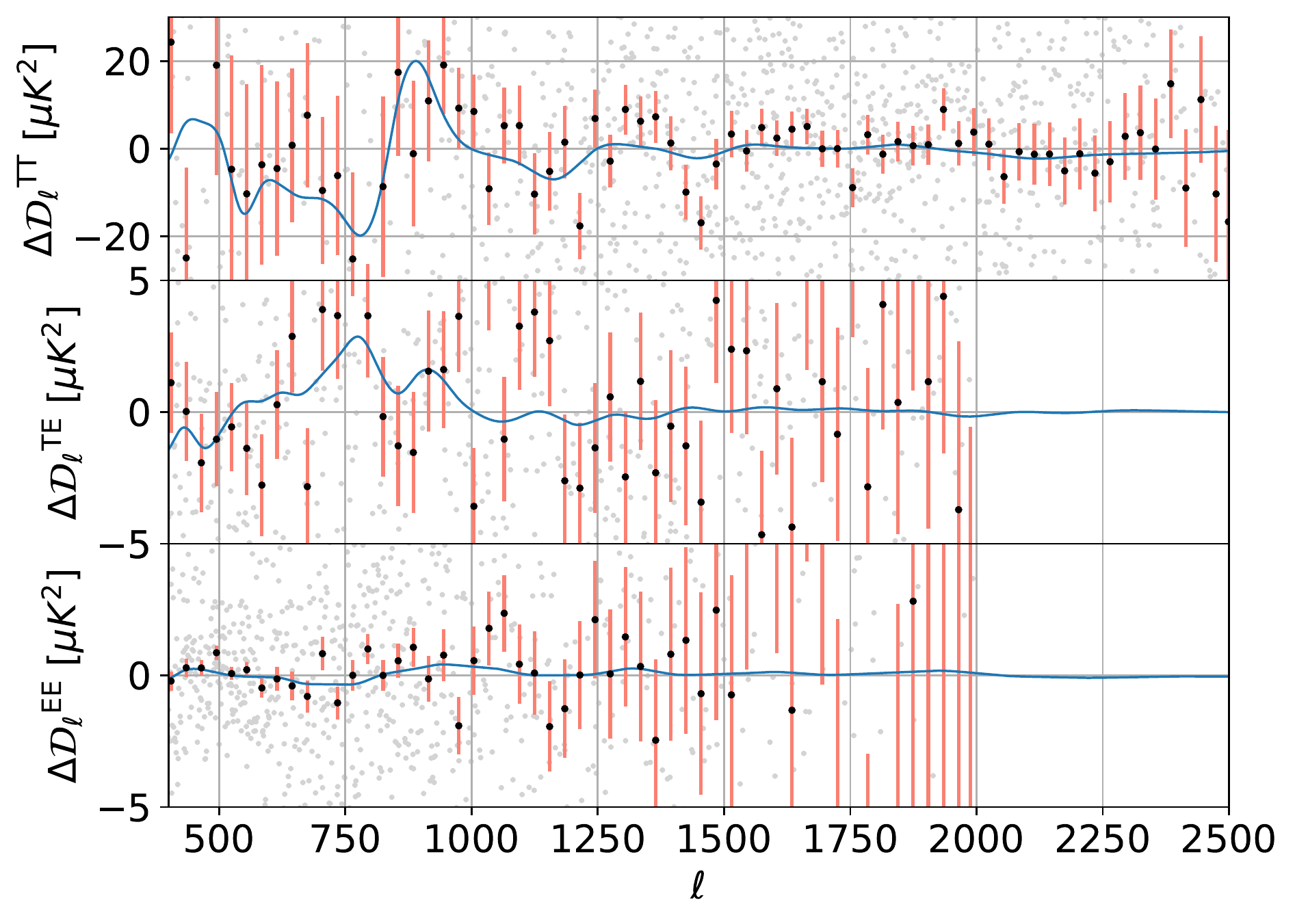}
    \caption{The difference between the GA best-fit of the unbinned data with the fast trigonometric functions and the Starobinsky prior for the $TT$, $TE$ and $EE$ spectra.
    The black points with light-red error bars correspond to the binned Planck 18 data, while the light-grey points for $\ell>30$ correspond to the unbinned ones.
    Note that for $\ell <30$, binned and unbinned data are the same.
    \textbf{Upper panel:} The full range of the data. The GA does not address the low-$\ell$ anomaly in the CMB.
    \textbf{Lower panel:} Zoom in the high-$\ell$ region, where the GA systematically improves the fit to CMB residuals.
    As can be seen in both regimes, the GA is never overfitting the unbinned data (which are the ones used in the analysis), but rather improving the overall fit.
    This is further confirmed in App.~\ref{sec:cva}.
    }
    \label{fig:delta_DL}
\end{figure*}

We now confirm \textit{a posteriori} the absence of overfitting in the best-fit candidate by looking at the residuals in the CMB spectra.
Specifically, in Fig.~\ref{fig:delta_DL} we plot the difference in the temperature and polarization spectra between the GA best-fit and the Starobinsky prior.
The black points with light-red error bars correspond to the binned Planck 18 data, while the light-grey points for $\ell>30$ are the unbinned ones.
Note that for $\ell < 30$, binned and unbinned data are the same.
For the GA best-fit, we do not bin the spectra, and rather show the full outcome of the algorithm as a blue solid line.
In the upper panel we show the full range of the data, while in the lower one we zoom in the high-$\ell$ region.
As can be seen in both panels, the GA is not noise-fitting, i.e. it is not trying to pass through every single (unbinned) data point and overfitting the data.
Rather, it is systematically improving the fit to the data, as can be seen by comparing it to the binned data (not used in the analysis).

In particular, no strongly oscillating behaviour is observed.
This is true for the $TT$, $TE$ and $EE$ spectra in both the low-$\ell$ and high-$\ell$ regimes.
We also note that the GA best-fit does not address the low-$\ell$ anomaly in the CMB, but rather improves the fit to high-$\ell$ residuals missed by the slow-roll, featureless, prior.
In Appendix~\ref{sec:cva} we perform additional statistical tests to quantitatively assess how much systematic is the fit of the GA to the data.

\section{Conclusions \label{sec:conclusions}}

In this work we present the proof of concept that a machine learning approach, based on genetic algorithms, can be used to reconstruct the inflationary potential directly from the CMB data.
In our set-up, we restrict to single-field inflation and use the Planck 18 CMB data to explore the functional space of inflationary potentials.
Since our pipeline is very modular, this work could be extended to more general inflationary scenarios, including multifield inflation or theories with higher-order derivatives.
Moreover, other CMB data sets could be added.

Our main results are twofold.
First, we show how our approach can reconstruct featureless slow-roll potentials, starting from a variety of different physical priors: quadratic, hilltop or Starobinsky inflation, and with different grammars -- polynomial or trigonometric (slow).
As expected, we find that the GA easily beats the quadratic inflation prior ($\Delta \chi^2=-3.175$ for $\alpha=0.4$ and $\Delta \chi^2=-4.538$ for $\alpha=1$, where $\alpha$ parameterizes the size of the steps in functional space).
We also confirm the robustness of the other two priors, Starobinsky and hilltop, where the improvements from the GA are marginal, $\Delta \chi^2\sim-0.1$.

After these consistency checks, we apply for the first time our machine learning approach to go beyond slow-roll and search for oscillatory features in the inflationary potential.
In this case, we use the unbinned Planck data and we allow for ``fast'' trigonometric functions in the grammar, i.e. containing small wavelengths in $\phi$-space.
Starting from the Starobinsky prior again, the GA now finds a dramatic improvement of $\Delta \chi^2=-21.570$ with respect to the prior.
To our knowledge, this is the best improvement to date upon slow-roll inflation when reconstructing directly the inflationary potential with latest Planck data.

Interestingly, the best-fit candidate with features contains a multimodal distribution of frequencies in field space, whose inverses can be interpreted as axion decay constants of the order of $f\sim 10^{-3}\,M_\mathrm{Pl}$.
These oscillatory modes in the potential also induce oscillations on the primordial power spectrum $P_\zeta(k)$ that are linear in $\log k$-space, and with a distribution of frequencies peaked at values of approximately $\omega_{\log}/H\sim [5, 9, 15, 30 ]$.
A statistical analysis of the $68.3\%$ best of all functions at the last generation proposed by the GA shows that the distribution of axion decay constants is not noisy and is clustered around specific values, showing that some frequencies in field space are systematically improving the fit to the Planck data.

A common concern of machine-learning approaches is that best-fit candidates proposed by the algorithm could be overfitting the data.
In our approach, overfitting is mitigated \textit{a priori} by the inclusion of a maximal length for the potentials proposed by the algorithm.
The absence of important overfitting is also checked \textit{a posteriori}, as can be seen in the plots of the temperature and polarization spectra shown in Fig.~\ref{fig:delta_DL}, and further confirmed in App.~\ref{sec:cva} with a dedicated analysis.
Finally, it ought to be said that the features fitted by the GA could in principle be attributed to unknown systematics in Planck likelihood, although we are not aware of systematics of this form.

In conclusion, we find that our machine-learning approach is capable of improving upon a given prior by proposing new inflationary potentials that are better fitted to data.
We show a first concrete application that demonstrates the potential of the GA to search for features in the data.
Our findings motivate the search for resonant features in the primordial power spectrum with a multimodal distributions of frequencies.

\textit{Numerical Analysis Files}: The \texttt{python} codes used by the authors in the analysis of this paper will be made publicly available upon publication at \href{https://github.com/snesseris/GA-inflation}{https://github.com/snesseris/GA-inflation}  \\

\section*{Acknowledgements}
The authors would especially like to thank
M.~Braglia for many useful discussions on the topic of this work. They are also grateful to 
G.~Herdoiza,
S.~Kuroyanagi,
M.~Martinelli and 
M.~de los Rios 
for useful discussions, and they acknowledge support from the research projects PGC2018-094773-B-C32 and PID2021-123012NB-C43, and the Spanish Research Agency (Agencia Estatal de Investigaci\'on) through the Grant IFT Centro de Excelencia Severo Ochoa No CEX2020-001007-S, funded by MCIN/AEI/10.13039/501100011033. \\
LP would also like to acknowledge support from the “Atracción de Talento” grant 2019-T1/TIC15784.  \\
Finally, the authors also acknowledge use of the ``San Calisto'' supercomputer at the IFT (courtesy of M.~Martinelli), use of the IFT Hydra cluster, use of the supercomputer at the ``Centro de Supercomputaci\'on de Galicia" and the packages \texttt{CLASS} and \texttt{clik}.

\begin{figure*}[!t]
\centering
\includegraphics[width=0.495\textwidth]{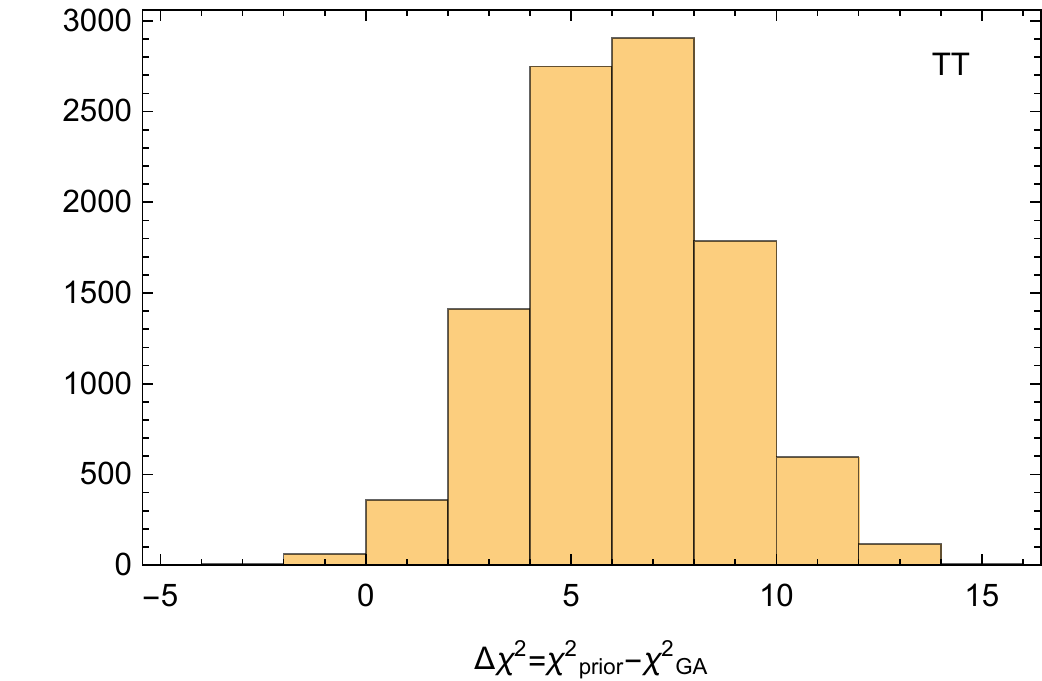}
\includegraphics[width=0.495\textwidth]{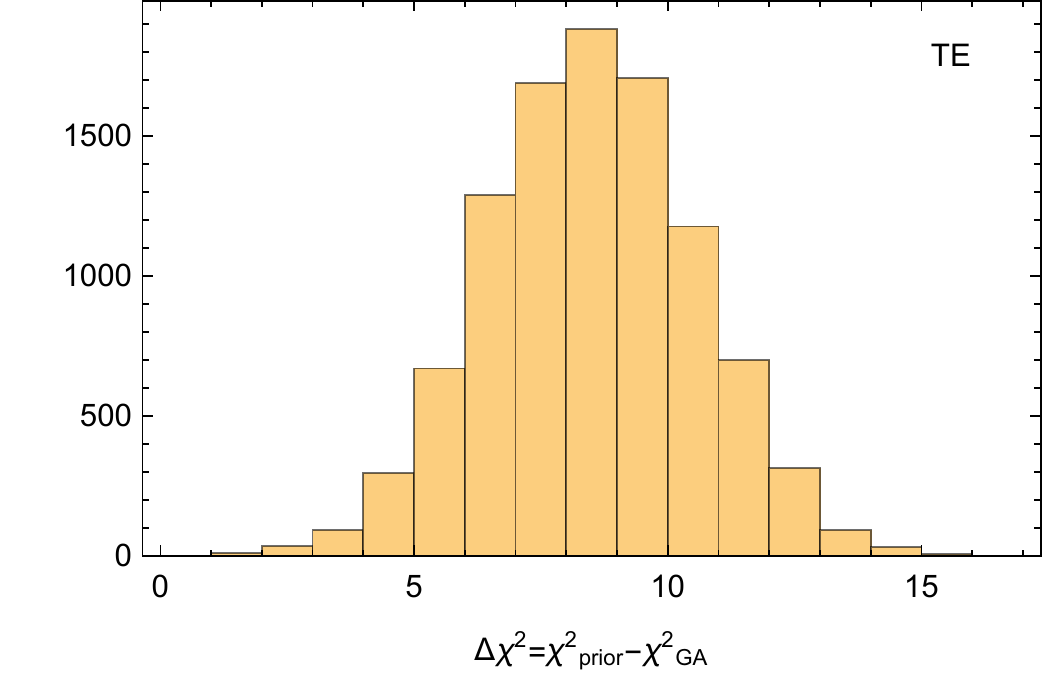}
\includegraphics[width=0.495\textwidth]{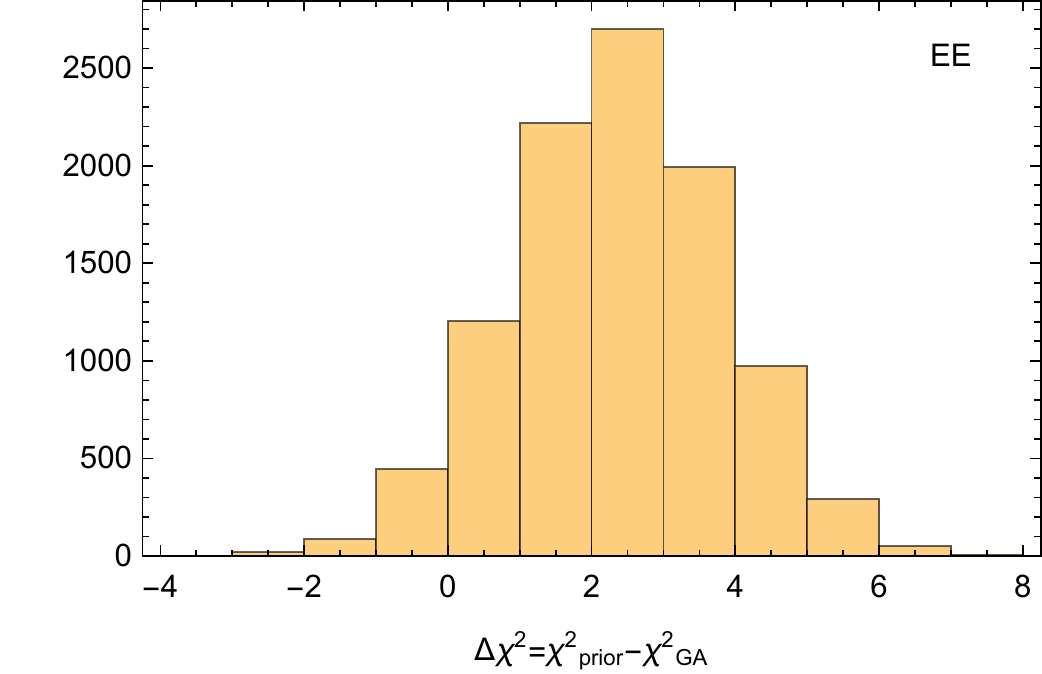}
\includegraphics[width=0.495\textwidth]{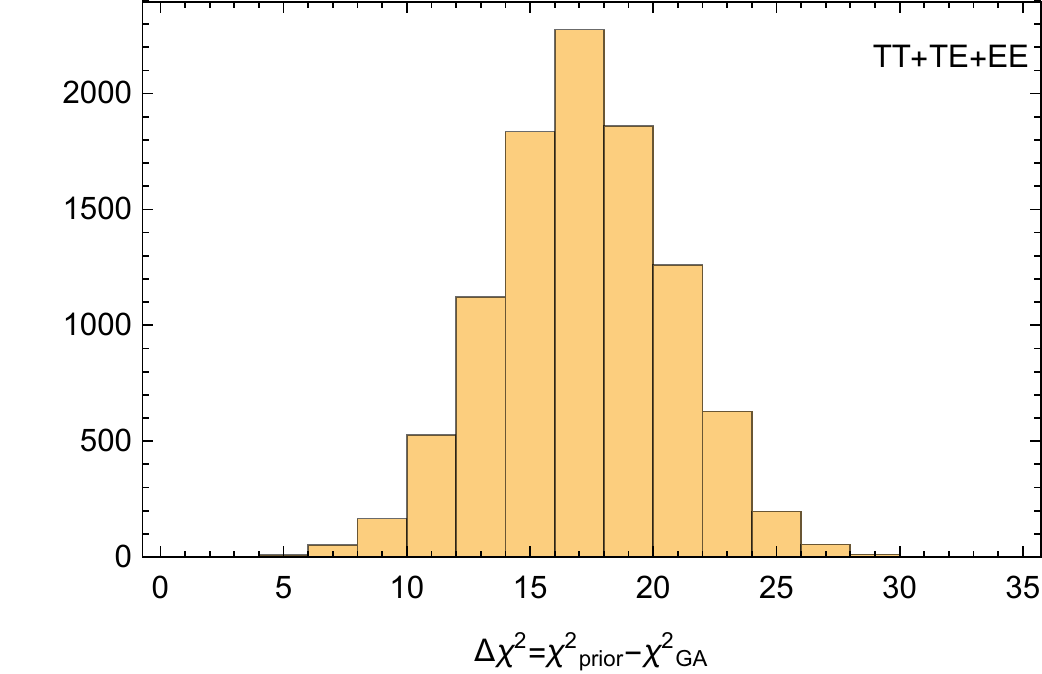}
\caption{Distribution of the $\chi^2$ values given by the simplified expression of Eq.~\eqref{eq:chi2_cva} for 10,000 realizations, where randomly $20\%$ of the unbinned Planck CMB data points have been removed. We show the $TT$, $TE$, $EE$ and $TT+TE+EE$ combinations and as can be seen, the GA consistently outperforms the prior, thus reinforcing our conclusion that no overfitting is taking place according to the cross-validation criterion.}
\label{fig:histograms_chi2}
\end{figure*}

\appendix
\section{Cross-validation checks\label{sec:cva}}

Here we summarize our \textit{a posteriori} checks to ensure the GA is not overfitting, using the approach of cross-validation (see Ref.~\cite{Peiris:2009wp} for a similar analysis with CMB data).

Specifically, cross-validation consists of comparing how well two models fit many different sub-samples of the same data, and checking the one better fitting the full data set is also consistently better fitting (statistically speaking) the sub-samples of it.
However, in practice, we cannot take sub-samples of the CMB data using the official Planck likelihood, as the latter does not simply consist of data points with error bars.
Modifying the Planck likelihood to accommodate for cross-validation checks is certainly an interesting direction for future work, but also clearly beyond the scope of this work.
Therefore, instead we build a simplified chi-squared-based likelihood using the publicly available CMB angular power spectra. \footnote{These official CMB power spectra $C_\ell^{XX}$ can be found at the Planck Legacy Archive: \url{https://pla.esac.esa.int/\#cosmology} }

Using the CMB spectra $C_\ell^{XX}$ we create a simplified chi-squared as:
\be 
\chi^2_{XX}=\sum_{\ell\in \mathrm {set}}\left[\frac{C_\ell^{XX,\mathrm{Planck}}-C_\ell^{XX,\mathrm{GA}}}{\sigma_{C_\ell^{XX}}}\right]^2,\label{eq:chi2_cva}
\ee
where $XX=(TT,TE,EE)$, we have ignored the covariance between the spectra and ``set" is a randomly chosen sub-sample of the full data set.
For our cross-validation tests, we keep only $80\%$ of the total data points in multipole space in the $TT$, $TE$ and $EE$ spectra respectively.
In what follows, we consider separately all the spectra, but also as a combination with a total chi-squared $\chi^2=\chi^2_{TT}+\chi^2_{TE}+\chi^2_{EE}$.

Here we should mention some caveats: first, this simplified likelihood is not suitable for proper statistical analyses of the Planck CMB data as it lacks the covariances between the spectra, but also pixel-level corrections at low multipoles. We only use it here a proxy for the full Planck likelihood, as the cross-validation test cannot be done with the latter, due to its complexity. Indeed, modifying the Planck likelihood itself is beyond the scope of this work. Another consequence of not being able to perform the cross-validation analysis with the true Planck data, is that we can only do it \textit{a posteriori} with the best-fit obtained with the true data, as otherwise the GA would be wrongly using our much simplified likelihood.

Having defined the random samples and the $\chi^2$, then we calculate the difference in the $\chi^2$ between the GA best-fit and the Starobinsky prior for 10,000 realizations of having removed $20\%$ of the data in random multipoles $\ell$. If the GA is overfitting, e.g. by passing through every point in some multipoles, then by removing $20\%$ of the points randomly a high enough number or times  we should find that the GA no longer outperforms the prior.

Performing this exercise we find the statistical distributions of the $\Delta \chi^2\equiv \chi^2_\mathrm{prior}-\chi^2_\mathrm{GA}$ for the different realizations and all three spectra. In Fig.~\ref{fig:histograms_chi2} we show the distribution of the $\Delta \chi^2$ values given by the simplified expression of Eq.~\eqref{eq:chi2_cva}. The different panels show the $TT$, $TE$, $EE$ and $TT+TE+EE$ combinations.
As can be seen, the GA best-fit consistently outperforms the prior in all three spectra, thus reinforcing our conclusion that no important overfitting is present in the best-fit.
In particular, we find that not a single realization over the 10,000 sub-data sets, is not better fitted by the GA best-fit than by the prior, when considering the total $\Delta \chi^2$ summing over all three spectra.

Quantitatively, we find that in the case of the $TT$ spectra the mean $\mu_{\Delta \chi^2}$ and standard deviation $\sigma_{\Delta \chi^2}$ of the $\Delta \chi^2$ are $(\mu_{\Delta \chi^2},\sigma_{\Delta \chi^2})=(6.30,2.54)$ respectively, for the $TE$ they are $(\mu_{\Delta \chi^2},\sigma_{\Delta \chi^2})=(8.49,2.06)$, for the $EE$ $(\mu_{\Delta \chi^2},\sigma_{\Delta \chi^2})=(2.36,1.47)$ and for the total $TT+TE+EE$ they are $(\mu_{\Delta \chi^2},\sigma_{\Delta \chi^2})=(17.15,3.59)$. As can be seen from the plots in Fig.~\ref{fig:histograms_chi2} and the corresponding mean values, there is statistically significant evidence that the GA best-fit consistently outperforms the prior by a wide margin, whatever chunk of the data being removed, thus assessing its robustness.

\bibliography{biblio}

\end{document}